\documentclass[fleqn,usenatbib]{mnras}

\usepackage{newtxtext,newtxmath}
\usepackage[T1]{fontenc}

\DeclareRobustCommand{\VAN}[3]{#2}
\let\VANthebibliography\thebibliography
\def\thebibliography{\DeclareRobustCommand{\VAN}[3]{##3}\VANthebibliography}

\usepackage{graphicx}	% Including figure files
\usepackage{amsmath}	% Advanced maths commands
\usepackage{subfig}
\newcommand{\frb}{FRB~20220912A}
\newcommand{\rone}{FRB~20121102A}
\newcommand{\rthree}{FRB~20180916B}
\newcommand{\rmeightyone}{FRB~20200120E}
\newcommand{\rsixtyseven}{FRB~20201124A}
\newcommand{\ronetwin}{FRB~20190520B}
\title[Forests of Microshots in FRBs]{Dense Forests of Microshots in Bursts from FRB~20220912A}

\author[D. M. Hewitt et al.]{
Dant\'e M. Hewitt,$^{1}$\thanks{E-mail: d.m.hewitt@uva.nl}
Jason W. T. Hessels,$^{1,2}$
Omar S. Ould-Boukattine,$^{2,1}$
Pragya Chawla,$^{1}$
Isma\"el Cognard,$^{3,4}$ \newauthor
Akshatha Gopinath,$^{1}$
Lucas Guillemot,$^{3,4}$
Daniela Huppenkothen,$^{5,1}$
Kenzie Nimmo$^{6}$
and
Mark P. Snelders$^{2,1}$
\\
$^{1}$Anton Pannekoek Institute for Astronomy, University of Amsterdam, Science Park 904, 1098 XH, Amsterdam, The Netherlands\\
$^{2}$ASTRON, Netherlands Institute for Radio Astronomy, Oude Hoogeveensedijk 4, 7991 PD Dwingeloo, The Netherlands\\
$^{3}$Station de Radioastronomie de Nan\c{c}ay, Observatoire de Paris, PSL University, CNRS, Universit\'e d'Orl\'eans, F-18330 Nan\c{c}ay, France\\
$^{4}$Laboratoire de Physique et Chimie de l'Environnement et de l'Espace LPC2E UMR7328, Universit\'e d'Orl\'eans, CNRS, F-45071 Orl\'eans, France\\
$^{5}$SRON Netherlands Institute for Space Research, Niels Bohrweg 4, 2333 CA, Leiden, Netherlands\\
$^{6}$MIT Kavli Institute for Astrophysics and Space Research, Massachusetts Institute of Technology, 77 Massachusetts Ave, Cambridge, MA 02139, USA
}

\date{Accepted XXX. Received YYY; in original form ZZZ}

\pubyear{2015}

\begin{document}
\label{firstpage}
\pagerange{\pageref{firstpage}--\pageref{lastpage}}
\maketitle

\begin{abstract}
We report on exceptionally bright bursts ($>$400\,Jy\,ms) detected from the repeating fast radio burst source \frb\ using the Nan\c{c}ay Radio Telescope (NRT), as part of the ECLAT (Extragalactic Coherent Light from Astrophysical Transients) monitoring campaign. These bursts exhibit extremely luminous, broadband, short-duration structures ($\sim16$\,$\mu$s), which we term `microshots' and which can be especially well studied in the NRT data given the excellent signal-to-noise and dynamic range (32-bit samples). The estimated peak flux density of the brightest microshot is 450\,Jy. We show that the microshots are clustered into dense `forests', by modelling them as Weibull distributions and obtaining Weibull shape parameters of approximately 0.5. Our polarimetric analysis reveals that the  bursts are nearly 100\,\% linearly polarised; have $\lesssim 10$\,\% circular polarisation fractions; a near-zero average rotation measure of 0.10(6)\,rad\,m$^{-2}$; and varying polarisation position angles over the burst duration. For one of the bursts, we analyse raw voltage data from simultaneous observations with the Westerbork RT-1 single 25-m dish. These data allow us to measure the scintillation bandwidth, 0.30(3)\,MHz, and to probe the bursts on (sub-)microsecond timescales. Some important nuances related to dedispersion are also discussed. We propose that the emission mechanism for the broadband microshots is potentially different from the emission mechanism of the broader burst components  which still show a residual drift of a few hundred MHz\,ms$^{-1}$ after correcting for dispersion using the microshots. We discuss how the observed emission is phenomenologically analogous to different types of radio bursts from the Sun.
\end{abstract}


%%%%%%%%%%%%%%%%%%%%%%%%%%%%%%%%%%%%%%%%%%%%%%%%%%

%%%%%%%%%%%%%%%%% BODY OF PAPER %%%%%%%%%%%%%%%%%%

\section{Introduction} \label{sec:intro}

Fast radio bursts (FRBs) are coherent radio transients, typically lasting a few milliseconds. Their extragalactic origin was inferred from their excess dispersion measures (DMs) and has been confirmed by multiple localisations to a diverse set of host galaxies \citep[for a recent review of FRBs, see][]{petroff_2022_aarv}. A small fraction \citep[$\sim2.6$\,\%;][]{chime_2023_apj} of known FRBs are confirmed repeaters \citep{spitler_2016_natur} that emit multiple bursts --- more often than not, sporadically and infrequently.  These repeater bursts are, on average, longer in duration and narrower in observed emission bandwidth compared to the apparently non-repeating FRBs \citep{pleunis_2021_apj}. Conversely, it has recently been demonstrated that the repetition rates of repeaters and the upper limits on repetition from apparently non-repeaters are not clearly distinct, after exposure and sensitivity corrections \citep{chime_2023_apj}. There are, however, some anomalously prolific repeaters that have appeared to suddenly awaken from quiescence, e.g., \rsixtyseven\ \citep{chime_2021_atel}. A non-negligible fraction of apparently non-repeating sources may thus in fact be repeaters (either with low repetition rates or long periods of quiescence); if so, their shorter-duration and wider-bandwidth bursts could arise from a different emission mechanism and/or beaming effects \citep{connor_2020_mnras_497}.

Many theories advocate for highly magnetized neutron stars as the progenitors of FRBs. This hypothesis has been strengthened by the detection of an extremely bright, millisecond-duration radio burst from the Galactic magnetar SGR~1935+2154 \citep{bochenek_2020_natur,chime_2020_natur_galacticfrb}. It is presently unclear whether the repeating and apparently non-repeating FRBs originate from the same progenitor(s). So far, there also seems to be no distinction in the properties of the host galaxies of repeaters and apparent non-repeaters \citep{bhandari_2022_aj,gordon_2023_arxiv}. If assuming a universal magnetar progenitor, however, different formation channels are required in order to reconcile the rate of all extragalactic FRBs with the volumetric population of magnetars \citep{margalit_2020_apjl,kirsten_2022_natur}. 

The repeater population itself, also boasts rich diversity. Repeaters have been localised to various types of host galaxies, including dwarf galaxies \citep[e.g., \rone\ and \ronetwin,][]{chatterjee_2017_natur,niu_2022_natur}, spiral galaxies \citep[e.g., \rthree,][]{marcote_2020_natur} and even a globular cluster in the nearby M81 galactic system \citep{kirsten_2022_natur}. Furthermore, the propagation effects we can measure from bursts such as scattering times, Faraday rotation measure (RM), and DM (as well as the evolution of these properties) suggest diverse, yet quite often chaotic, magneto-ionic local environments for the repeaters \citep[e.g.,][]{michilli_2018_natur,mckinven_2023_apj}.  There are various other characteristics that can be used to group repeaters: \rone\ and \ronetwin\ are the only repeaters so far associated with a persistent radio source \citep{chatterjee_2017_natur,marcote_2017_apjl,niu_2022_natur}, potentially a nebula powered by the FRB source; \rone\ and \rthree\ are the only repeaters so far that show an underlying periodicity in their activity windows  \citep[of $\sim16$ and $\sim160$ days, respectively;][]{chime_2020_natur_periodicactivity,rajwade_2020_mnras,cruces_2021_mnras}; \rthree\ and \rmeightyone\ are the only repeaters so far that exhibit (sub-)microsecond burst structure \citep{nimmo_2021_natas,nimmo_2022_natas}; and only a handful of repeaters are known to enter periods of high activity \citep[`burst storms', e.g.,][]{li_2021_natur,hewitt_2022_mnras,zhou_2022_raa,nimmo_2023_mnras}. It is thus also unclear whether all repeaters originate from the same progenitor(s). The relationship between the various FRB sources we observe is clearly intricate, but detailed studies of burst properties can help further our understanding.

On 15 October 2022, the CHIME/FRB Collaboration announced that a new repeating FRB source, \frb\, had become active. Within 3 days (with a daily exposure time of only $\approx10-15$\,minutes) CHIME/FRB detected 9 bursts from this source at $400-800$\,MHz and report a DM of 219.46\,pc\,cm$^{-3}$ \citep{mckinven_2022_atel}. \frb\ has since proven to be one of the most active known repeaters to date. The vast majority of the bursts have been detected around $\sim1.4$\,GHz (L-band), where multiple telescopes, including the 100-m Green Bank Telescope (GBT), Five hundred meter Aperture Spherical Telescope (FAST), and Effelsberg 100-m telescope have detected more than a hundred bursts within a few hours or less of observations \citep[e.g.,][]{feng_2022_atel,zhang_2022_atel,kirsten_2022_atel,feng_2023_arxiv,zhang_2023_arxiv}. Bursts have also been detected at lower frequencies  \citep[$\sim300-400$\,MHz;][]{bhusare_2022_atel,ouldboukattine_2022_atel_15817} and at higher frequencies around 2\,GHz \citep[S-band;][]{perera_2022_atel,rajwade_2022_atel}. Thus far, however, no bursts have been detected above 3\,GHz despite many hours of observations \citep{kirsten_2022_atel,rajwade_2022_atel,sheikh_2022_atel}, or at the operating frequencies of the LOFAR High-Band Antennas where other repeaters have been seen \citep[110-190\,MHz;][]{gopinath_2023_arxiv}. 

The majority of repeaters exhibit high fractions of linear polarisation, and little to no circular polarisation \citep[e.g.,][]{gajjar_2018_apj,day_2020_mnras,nimmo_2021_natas}. The brightest burst detected from \frb by CHIME/FRB ($\sim 600$\,MHz), also exhibited 100\% linear polarisation, and showed an RM value of\footnote{Note that, when convenient, we use parentheses to indicate the uncertainty on the least significant digits.} +0.6(1)\,rad\,m$^{-2}$ \citep{mckinven_2022_atel}.  Bursts detected at higher frequencies ($\sim1.4$\,GHz) from FAST and the GBT show similar properties: 100\% linear polarisation and relatively stable RM values close to zero rad\,m$^{-2}$, the latter being indicative of a clean local environment. Additionally, approximately half of the bursts have some degree of circular polarisation, making \frb\ the repeater with the highest known fraction of bursts that exhibit circular polarisation \citep{feng_2023_arxiv,zhang_2023_arxiv}. Furthermore, some bursts have polarisation position angle (PPA) swings, which is also scarce among the known repeaters \citep[see ][]{luo_2020_natur}.

\frb\ was localised to a host galaxy with redshift $z=0.0771(1)$ by the Deep Synoptic Array (DSA-110) during its commissioning phase \citep{ravi_2023_apjl}. The host, PSO~J347.2702+48.70, is a massive galaxy (log\,M$_*=10.0(1)$) with moderate star formation (SFR\,$\gtrsim0.1\,$M${_{\odot}}$\,yr$^{-1}$), not unlike other FRB host galaxies \citep{gordon_2023_arxiv}. The host galaxy is located at a luminosity distance of 362.4(1)\,Mpc and classified as composite in the Baldwin, Phillips \& Terlevich (BPT) scheme.

In this paper, we present three exceptionally bright bursts detected from \frb\ using the Nan\c{c}ay Radio Telescope (NRT). For two of the bursts, we have simultaneous coverage with the Westerbork RT-1 single 25-m dish, which recorded raw voltage data, allowing us to study burst properties at the Nyquist limit, but with lower bandwidth and sensitivity compared to NRT. The excellent quality of these combined data, together with the extraordinary signal-to-noise (S/N) of the bursts, provide a great opportunity for an in-depth study of burst features. In Section~\ref{sec:observations} we describe the observational setup and data acquisition. In Section~\ref{sec:analysis} we describe our search for bursts and the analysis used to investigate the burst properties. In Section~\ref{sec:discussion} we discuss the challenges related to determining an accurate DM. This is critical for the subsequent discussion where we probe the range of timescales of the emission in these bursts from \frb\ and provide observational evidence that FRBs can contain short timescale (and sometimes narrow-band) shots of emission that are tightly packed together. We then compare our findings with what is seen from other FRBs, pulsar giant pulses, magnetar bursts, and solar radio bursts. Finally, in Section~\ref{sec:conclusions} we conclude and put forward some recommendations for future work. Throughout the paper, unless otherwise mentioned, numbers in parentheses of our measurements are used to indicate the 3$\sigma$-uncertainty on the least significant digits.

\section{Observations} \label{sec:observations}

\subsection{Nan\c{c}ay Radio Telescope}

The NRT is a Kraus type meridian telescope situated in the centre of France. At $\sim1.4$\,GHz the NRT has a system temperature of $T_{\rm sys}\approx35$\,K and a gain of $G\approx1.4$\,K/Jy, making it effectively as sensitive as a 100-m radio dish. At the beginning of 2022 we established a monitoring campaign called ECLAT\footnote{{\it \'{E}clat} can be translated as `burst' in French.} (Extragalactic Coherent Light from Astrophysical Transients; PI: D.~M.~Hewitt) on the NRT. ECLAT is conducting follow-up observations of about a dozen repeaters with approximate weekly cadence over the course of two years. Thus far, bursts have been detected from more than half of the sources, establishing the NRT as a valuable instrument for FRB research. Following the detection of \frb\ by CHIME/FRB we included this source among the repeaters that are regularly monitored. The pointing position we used was that of the original DSA-110 localisation: RA=23$^{\rm{h}}$09$^{\rm{m}}$05.49$^{\rm{s}}$ Dec=+48$^{\circ}$42$^{\prime}$25.6$^{\prime\prime}$ \cite[J2000;][]{ravi_2022_atel}. The offset from the final localization: RA=23$^{\rm{h}}$09$^{\rm{m}}$04.9$^{\rm{s}}$ Dec=+48$^{\circ}$42$^{\prime}$25.4$^{\prime\prime}$ \cite[J2000;][]{ravi_2023_apjl} is  $5.8^{\prime\prime}$ but this does not impact our observations as the NRT has a half-power beam width of $4^{\prime}$ (RA) $\times$ $22^{\prime}$ (Dec) at 1.4\,GHz.

These data were acquired using the Low Frequency receiver ($1.1-1.8$\,GHz) of the focal plane and receivers systems called FORT ({\it Foyer Optimis\'{e} pour le Radio T\'{e}lescope}), and have 512\,MHz of bandwidth (consisting of 128 channels) centered at a frequency of 1484\,MHz. The data were recorded as 8 subbands, each consisting of sixteen 4-MHz channels, using the Nan\c{c}ay Ultimate Pulsar Processing Instrument \citep[NUPPI;][]{desvignes_2011_aipc} with a time resolution of 16\,$\mu$s, 32-bit sampling (which is better than most FRB observations and hence provides excellent dynamic range for studying these bursts) and full polarisation information in a linear basis. Coherent dedispersion, within the 4-MHz channels, was applied using a DM of 219.46\,pc\,cm$^{-3}$ for \frb, as reported by \cite{mckinven_2022_atel}. The maximum DM offset that corresponds to a temporal smearing equal to the time resolution of the NRT data is 0.9\,pc\,cm$^{-3}$. We measure DMs for the detected bursts with a smaller offset than this and consequently, the residual temporal smearing due to inaccurate coherent dedispersion, even in the worst case scenario, is still less than half the NRT sampling time (DM determination is discussed further in Section~\ref{sec:DM_det}). Additionally, each observation is accompanied by an observation of a 3-Hz pulsed noise diode, which is used to calibrate the polarimetry. 

\subsection{Westerbork RT-1}

We were also monitoring \frb\ during this period using a single 25-m Westerbork dish (RT-1) in the Netherlands. The SEFD\footnote{\url{www.evlbi.org/sites/default/files/shared/EVNstatus.txt}} at 1.4\,GHz is 420\,Jy.  The Westerbork observations were conducted with an observing bandwidth of 128\,MHz centered at 1271\,MHz. Raw voltage data were recorded as 2-bit samples of both left and right circular polarisations with the local DBBC2 and Flexbuff systems, and stored in VLBI Data Interchange Format (VDIF; \citealt{whitney_2010_ivs}). Using \texttt{digifil} from the standard pulsar
software package \texttt{DSPSR} \citep{vanstraten_2011_pasa}, we created Stokes~I filterbanks from the voltage data, with frequency and time resolutions of 62.5\,kHz and 64\,$\mu$s, respectively, which we then used to search for bursts. We also created filterbank data products from the raw voltages at the time of detected bursts, with different time/frequency resolution combinations, for other analyses, as will be explained in later sections.

\section{Analysis and Results} \label{sec:analysis}

\subsection{Burst search}

\subsubsection{The ECLAT burst search pipeline}
NUPPI records data as eight 64-MHz subbands, which we stitch together and write out as 32-bit Stokes~I filterbanks with 512-MHz bandwidth using custom Python scripts that incorporate functionality from the \texttt{YOUR} (Your Unified Reader) library \citep{Aggarwal2020}. These filterbanks are converted to 8-bit samples using \texttt{digifil} for compatibility with the \texttt{Heimdall} software used later in the pipeline. The 8-bit filterbanks are then processed in two-minute chunks using the \texttt{rfifind} tool from the \texttt{PRESTO} pulsar software suite \citep{ransom_2001_phdt}, to determine which channels are most contaminated by radio frequency interference (RFI). We flag only these channels and apply no flagging to individual time intervals.  The flagged data are then searched using \texttt{Heimdall}, where we set a S/N threshold of 7. We effectively searched for boxcar widths ranging from 16\,$\mu$s to 131\,ms within a DM range of $200-250\,$pc\,cm$^{-3}$ for \frb. Candidates from the \texttt{Heimdall} search are classified by the machine-learning classifier \texttt{FETCH} \citep{aggarwal_2020_rnaas}. We manually inspect all the candidates for which any of the \texttt{FETCH} models A to H give a score above 0.5. If the bursts pass this manual inspection and are considered astrophysical, they are extracted from the original 32-bit data with full polarisation information. The 32-bit data are used for the remainder of the analyses.

We detected many hundreds of bursts from \frb, and while the burst rate drastically decreased around the end of 2022, we are still occasionally detecting bursts at the time of writing this manuscript.  The entire burst sample and a study of the evolution of the burst properties over time  will presented in a future paper. On 29 October 2022 (MJD~59881), 1 November 2022 (MJD~59884), and 16 November 2022 (MJD~59899), we detected exceptionally bright bursts from \frb, which we refer to as B1, B2 and B3, respectively, throughout the rest of this paper. These were the three bursts with the highest detection S/N. 

\begin{figure*}
    \centering
    \includegraphics[width=1\textwidth]{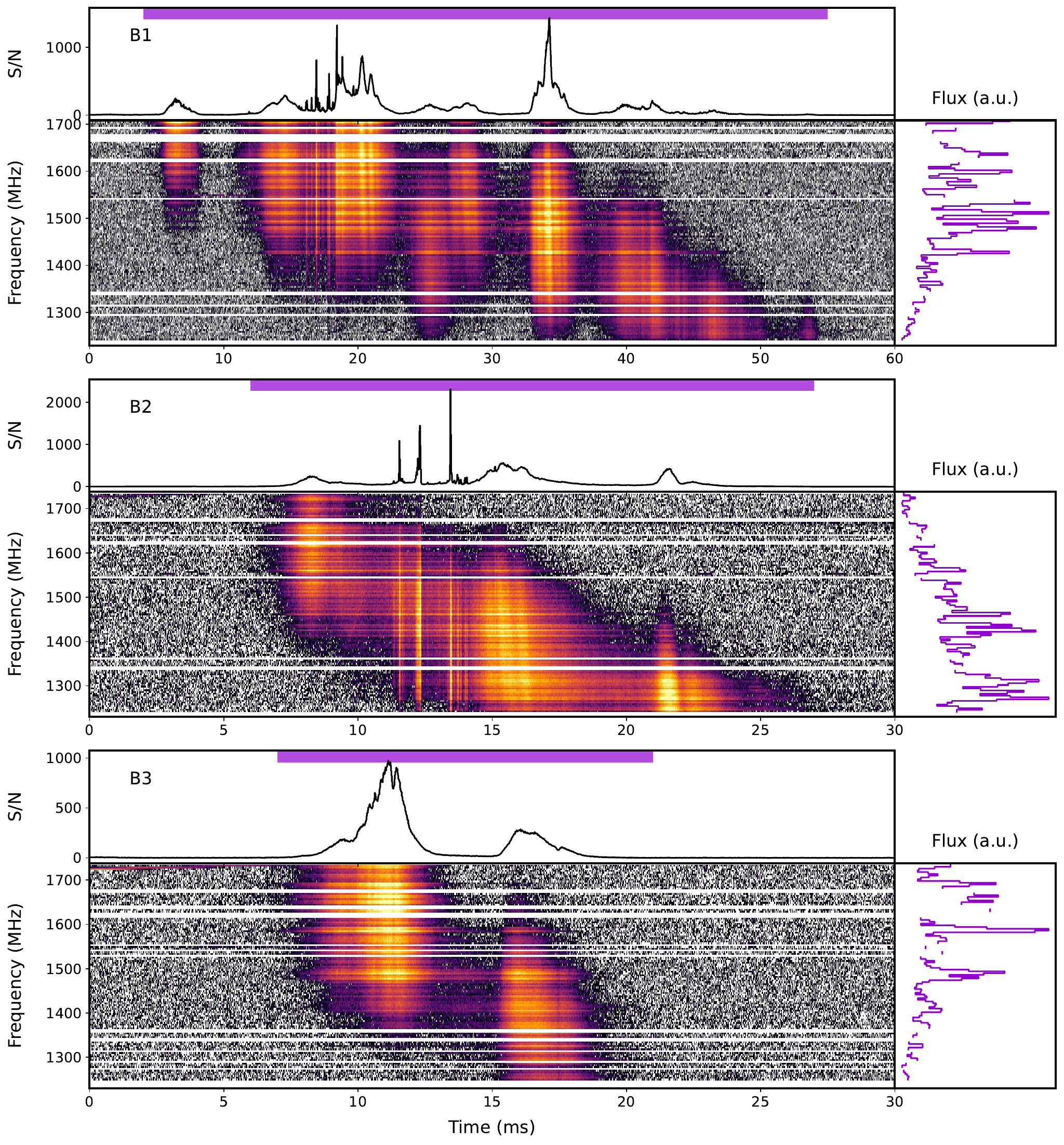}
    \caption{Dynamic spectra of three bursts detected from \frb\ with the NRT as part of the ECLAT monitoring campaign. The burst IDs (see Table~\ref{tab:properties}) are shown in the top-left corners of each sub-figure. The frequency and time resolution are 4\,MHz and 16\,$\mu$s, respectively. In addition to being coherently dedispersed to a DM of 219.46\,pc\,cm$^{-3}$, bursts B1, B2 and B3 have also been incoherently dedispersed to DMs of 219.356, 219.375 and 219.8\,pc\,cm$^{-3}$, respectively (see the main text for more details). Horizontal white lines in the dynamic spectra indicate frequency channels that have been excised due to the presence of RFI. The top panel of each sub-figure shows the frequency-averaged time series of the burst, and the right panel shows the time-averaged spectrum of the burst. The time series is obtained by summing over the entire observing bandwidth, while the spectrum is obtained by only summing over the times where the burst is present (determined by eye and indicated by the purple bar in the top panel). The spikes in the frequency spectra are likely due to scintillation in the Galactic ISM. Note that while the frequency extent of each dynamic spectrum is the same, the time extent is not. The noise is Gaussian-distributed with a standard deviation of 1 and a mean of 0. The colour scale for all sub-figures has been logarithmically scaled using the 99.9 quantile as the maximum and a value of 1 as the minimum.  Note that artifacts, arising from out-of-band emission before the first digitisation stage, are visible in the highest-frequency channels, preceding the burst.}
    \label{fig:family_plot}
\end{figure*}

In Figure~\ref{fig:family_plot}, we show the dynamic spectra, frequency-averaged burst profiles, and time-averaged spectra of the three brightest bursts detected thus far from \frb\ in our ECLAT observations. Each frequency channel in the dynamic spectrum has been individually normalised using off-burst statistics (i.e., bandpass corrected). The brightness variations in the spectra that are a few bins wide are likely the result of unresolved scintillation. The extreme S/N of these bursts results in some artifacts from out-of-band emission before the first digitisation stage that can be seen in the dynamic spectra at the highest observed frequencies, before the burst occurs.

\begin{figure*}
    \centering
    \includegraphics[width=\textwidth]{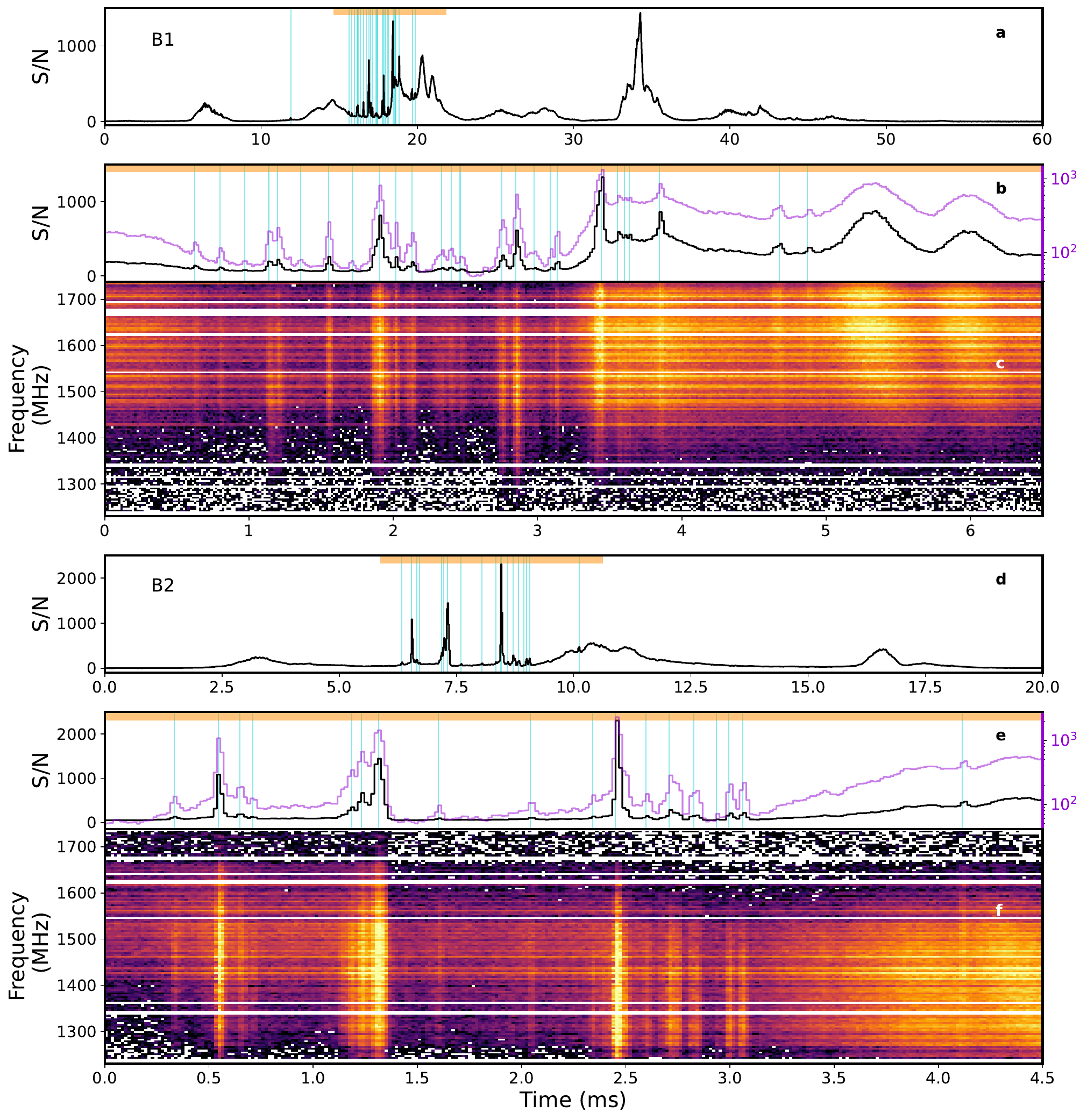}
    \caption{A zoomed-in view of bursts B1 and B2 is shown in the two sub-figures. Panels a and d show the full temporal burst profile of bursts B1 and B2, respectively. Microshots are indicated by vertical turquoise lines. In each of these panels, the time-span where the microshots are clustered is indicated by a horizontal orange bar. Panels b and e show zoom-ins of these regions. The purple profile is the same as the black, but on a logarithmic y-scale, revealing microshots that are less visible in the black linearly scaled profile due to the wide range of observed S/Ns. The dynamic spectra of these zoomed-in regions are shown in panels c and f. Horizontal white lines indicate channels that have been zapped due to the presence of RFI. Also note the residual frequency drift present in the broader burst components that occur after the microshot forests in panel c.}
    \label{fig:microshots}
\end{figure*}

Two of the three bright bursts exhibit clustered broadband (and sometimes exquisitely bright) emission components that are sometimes a single 16-$\mu$s time bin in duration. Throughout this paper we refer to these extremely short-timescale, broadband structures as `microshots'. In Figure~\ref{fig:microshots} the top panels show the time profile of the entire burst envelopes, while the middle and bottom panels display zoom-ins of the time profiles and dynamic spectra where the microshots appear. In the middle panels we have also over-plotted, in purple, the time profile on a logarithmic scale y-axis, which reveals many more microshots that are less apparent in the linear scale. The microshots have been manually identified and are indicated by vertical turquoise lines in the figure.

\subsubsection{Westerbork burst search}

The burst search pipeline used to search the Westerbork data has been previously described in detail \citep{kirsten_2021_natas,kirsten_2022_natur}. Westerbork was not on source during burst B1, but on 29 October 2022 and 16 November 2022 the ECLAT observations overlapped with Westerbork observations and B2 and B3 were independently discovered in the search of the Westerbork data. 

\subsection{Dispersion Measure determination}
\label{sec:DM_det}
We determined the DM of B1 using the three extremely bright and broadband microshots seen in the dynamic spectrum of the NRT data. We chose a DM that ensured that these microshots each individually arrive at all observed frequencies at the same time. This was done by dedispersing these three microshots to a range of trial DM values using a technique that shifts channels by fractions of a bin. We then fitted a Gaussian to these S/N versus DM curves to determine an optimal value and uncertainty. The microshots slightly differ from one another in their skewness, brightness, duration and bandwidth. Together with the intense variations in brightness as a function of frequency (likely scintillation, as we discuss later) this complicates the DM determination, but we attempt to mitigate this by averaging the power of a few microshots, and fitting a Gaussian to this averaged S/N vs DM curve. The resulting fits are shown in Appendix Figure~\ref{fig:sn_v_dm}. The curves are not smooth and exhibit scalloping that differs from one microshot to another. This occurs because the microshots have durations comparable to the time resolution of our data and our frequency resolution is rather coarse (since coherent dedispersion has been implemented beforehand). Artificially increasing the time resolution of the dynamic spectrum before dedispersion (with  nearest neighbour interpolation), and then downsampling to the original time resolution after the frequency channels have been shifted, results in slightly less scatter in S/N versus DM. In the Westerbork data, where sub-microsecond resolution is available, the scalloping disappears. Our fits yield a DM of 219.356(12)\,pc\,cm$^{-3}$ for B1. B2 similarly has three extremely bright and broadband microshots that were used to determine a DM of 219.377(9)\,pc\,cm$^{-3}$. Consistent measurements were obtained by performing similar analysis on the Westerbork data of B2 (time resolution of 1\,$\mu$s). B3, however, shows no microshots, even when exploring the burst with higher time resolution in the Westerbork data. Here we thus focused on straightening the notches (narrow intensity dips) that are present in the burst, resulting in a DM of 219.8(1)\,pc\,cm$^{-3}$. This is likely a slight overestimation of the DM because the DM that straightens the microshots in the other bursts does not necessarily straighten the dips in the broader sub-bursts. For the remainder of the analyses we use the aforementioned DM values, tabulated in Table~\ref{tab:properties}, unless stated otherwise.

In the literature, DMs are often determined using \texttt{DM\_phase} \citep{seymour_2019_ascl}\footnote{\url{https://github.com/danielemichilli/DM\_phase}}, which maximizes the coherent power of a burst across the bandwidth in order to determine a best DM. For the sake of comparison, using the entire burst as input for \texttt{DM\_phase}, while applying a bandpass filter to limit the fluctuation frequencies between 1 and 15.625\,ms$^{-1}$ (i.e., timescales ranging from 2 bins to 1\,ms) results in optimal DMs of 219.35(3), 219.37(2) and 219.8(2)\,pc\,cm$^{-3}$ for bursts B1, B2 and B3, respectively (1$\sigma$-uncertainties).  These values are consistent with those we obtain from aligning the microshots to arrive at the same time at all observed frequencies. All the aforementioned DM values are summarised in Table~\ref{tab:more_properties}. 

Moreover, these DMs are generally consistent with what was reported by CHIME/FRB and others thus far. In Section~\ref{subsec:dm} we further justify our decision for determining the DM in this manner and discuss other important nuances related to FRB DM determination in general.

\subsection{Measurement of burst properties}

Some properties of the bursts are tabulated in Table~\ref{tab:properties}. Before determining any properties we first removed the bandpass: for each channel we subtract the mean and divide by the standard deviation of the off-burst noise, free of any obvious RFI. To convert these pixels, which are now in S/N units, to physical flux densities we multiply by the radiometer equation. 

\begin{table*}
\caption{Burst properties}
\begin{center}
{\begin{tabular}{llrrrrr} \hline
Burst ID & TOA (MJD) $^a$  & Peak flux  & Fluence & Isotropic equivalent & Drift rate  & DM \\
 & & density (Jy)$^b$ & (Jy ms)$^{b,c}$ & energy (10$^{40}$\,erg)$^{b,d}   $ & (MHz\,ms$^{-1}$)$^e$ & (pc\,cm$^{-3}$)\\
\hline
B1           &    59881.86563025384     &    281    &  1089 & 8.1  & $-8$(3) & 219.356(12)  \\
B2           &    59884.83394049601     &    450    &  474    &  3.5 & $-20$(6)  & 219.377(9) \\
B3           &    59899.78465274653     &    190    &  411    & 3.1  & $-34$(8)  & 219.8(1) \\
\hline
\label{tab:properties}
\end{tabular}}\end{center}
$^a$ The topocentric time-of-arrival of the burst measured at the time bin of peak flux density, corrected to infinite frequency for a dispersion measure of 219.37\,pc\,cm$^{-3}$ and using a DM constant of 1/(2.41 $\times10^{-4}$)\, MHz$^2$\,pc$^{-1}$\,cm$^3$\,s.
\\
$^b$The estimated uncertainty is approximately 20\% due to uncertainty in the SEFD.
\\
$^c$ Determined over the full bandwidth of the observations.
\\
$^d$ Calculated as 4$\pi$F$\Delta\nu$D$_L^2$/(1+$z$), where F is the fluence, $\Delta\nu$ is the observing bandwidth, and  D$_L$ and $z$ are the luminosity distance and redshift of the host galaxy, respectively.
\\
$^e$ Calculated as cot$(\theta)$, where $\theta$ is the angle of the 2D Gaussian fit to the 2D ACF (see main text for details).
\\

\end{table*}

\subsubsection{ACF analysis}
We measured drift rates by calculating 2D autocorrelation functions (ACFs) of the bursts. We masked the zero-lag noise spike and then fitted a 2D Gaussian function to the ACF, following the technique detailed in \citet{chamma_2021_mnras}. Uncertainties on the drift rate are calculated as in \cite{gopinath_2023_arxiv}. The ACFs of the bursts can be seen in Appendix Figure~\ref{fig:nrt_acf}. The angle of the 2D Gaussian is defined as the orientation of the semi-major axis measured counterclockwise from the positive y-axis and the tangent of this angle gives the burst drift rates reported in Table~\ref{tab:properties}. These drift rates are comparable to drift rates measured in other complex bursts from FRBs such as FRB~20201124A \citep[e.g.,][]{zhou_2022_raa}, and with the lower-end of drifts measured for FRB~20121102A \citep[e.g.,][]{hessels_2019_apjl}. Measurement of drift rates at other radio frequency bands are necessary to determine whether the drift rate varies linearly with frequency for \frb\, as it does for other repeaters \citep[e.g.,][]{hessels_2019_apjl,gopinath_2023_arxiv}.

The 2D ACFs can further be used to measure the frequency and time scales of brightness variations in these bursts. The Galactic coordinates of \frb\ are $l = 106.1^{\circ}$, $b = -10.8^{\circ}$; it thus lies slightly outside of the Galactic plane and scatter broadening should consequently be moderate.  The expected scattering timescale for \frb\ at 1\,GHz according to the NE2001 Galactic electron density model \citep{cordes_2002_arxiv} is $2.6\,\mu$s. Assuming a frequency scaling of $\tau\propto\nu^{-4}$, the expected scattering time at the center of the Westerbork band (1.271\,GHz) is 1.0\,$\mu$s, corresponding to a scintillation bandwidth of 160\,kHz (using the approximation $\Delta\nu\sim1/2\pi\tau$). At 1.484\,GHz, which is the center of the NRT band, the scintillation bandwidth is expected to be approximately 290\,kHz, which is less than a tenth of the frequency resolution of these data (4\,MHz). Since the frequency resolution of our NRT data is too coarse to resolve the expected scintillation bandwidth we do not attempt to measure it in the frequency ACF. The presence of a one-bin peak at the centre of the frequency ACF, does however hint at a scintillation bandwidth of a few MHz at most. Fortunately, as previously mentioned, voltage data were recorded for B2 with Westerbork RT-1 and we were thus able to generate coherently dedispersed total intensity filterbanks with even higher time and/or frequency resolution than that of the NRT data, using the Super FX Correlator, \texttt{SFXC} \citep{keimpema_2015_exa}. We constructed data products with a frequency resolution of 62.5\,kHz and computed an ACF of these data, allowing us to distinguish between the different frequency scales that are present. We then fit a Lorentzian function to the narrow central feature present in the 1D frequency ACF and measure a scintillation bandwidth, defined as the half-width at half-maximum of the Lorentzian, of 300(30)\,kHz (see Figure~\ref{fig:wb_acf}). This corresponds to a scattering timescale of $\sim$0.5\,$\mu$s at 1.271\,GHz, which is consistent within a factor of 2 with the prediction from NE2001. Moreover, once frequency scaling has been taken into account, our measurement is also consistent with the scintillation bandwidth of 390(20)\,kHz reported from FAST observations that were conducted at a central frequency of 1.3\,GHz \citep{wu_2023_arxiv}.

\subsubsection{Power spectrum analysis}
In \cite{nimmo_2021_natas,nimmo_2022_natas} the power spectra of bursts are computed to determine if the bursts are consistent with red noise and to explore whether there are any significant features present in the power spectra such as quasi-periodic oscillations (QPOs). Following this same procedure and using the \texttt{Stingray} modelling interface \citep{huppenkothen_2019_apj}, we computed the power spectra of bursts B1 to B3 using the NRT data, as well as B2 with the Westerbork data (which has a 16$\times$ higher time resolution of 1\,$\mu$s), and modelled them as power laws (red noise), where the maximum-a-posteriori model is of the form: 

    \begin{equation}
        f(\omega)=A\omega^\alpha + C
    \end{equation}

\begin{figure*}
    \centering
      \subfloat[][]{\includegraphics[width=.48\textwidth]{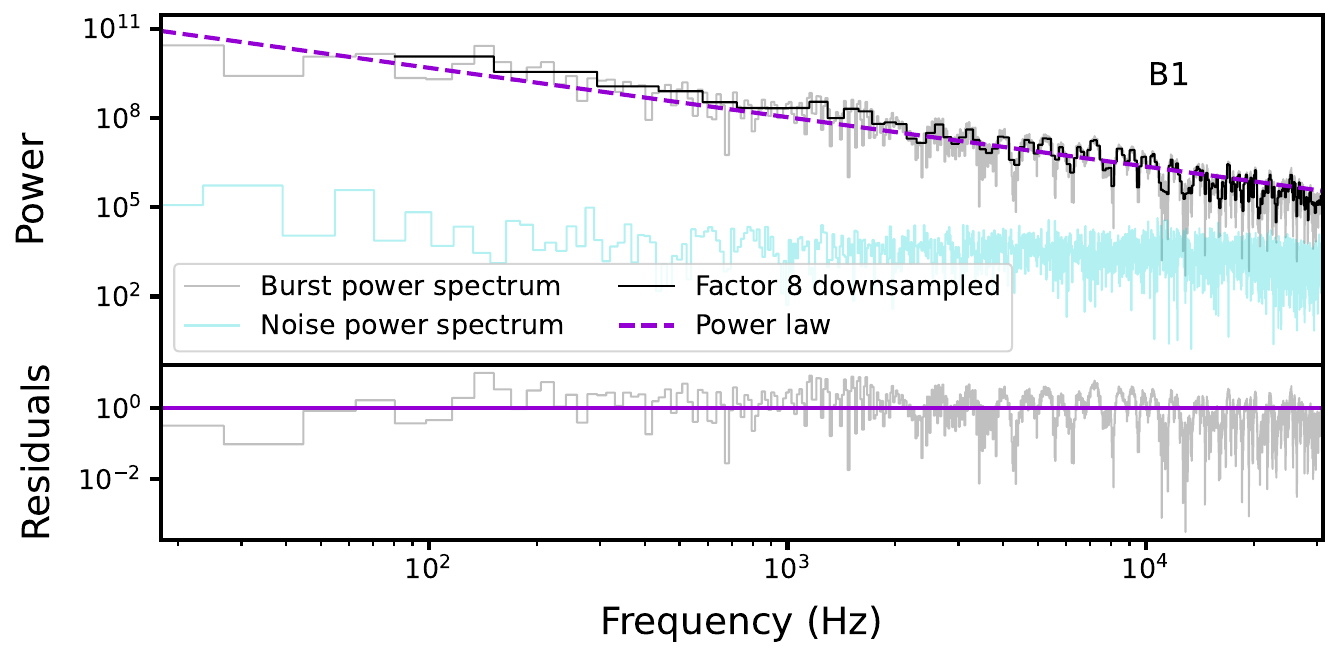}}\quad
      \subfloat[][]{\includegraphics[width=.48\textwidth]{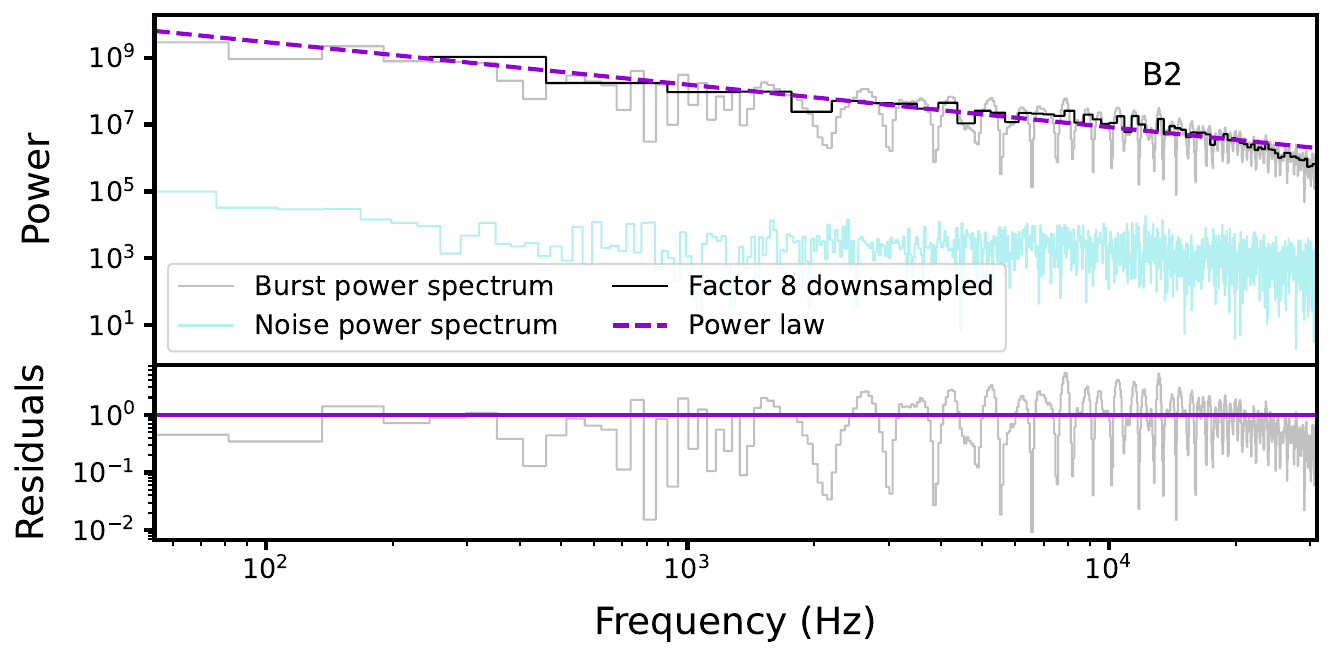}}\\
      \subfloat[][]{\includegraphics[width=.48\textwidth]{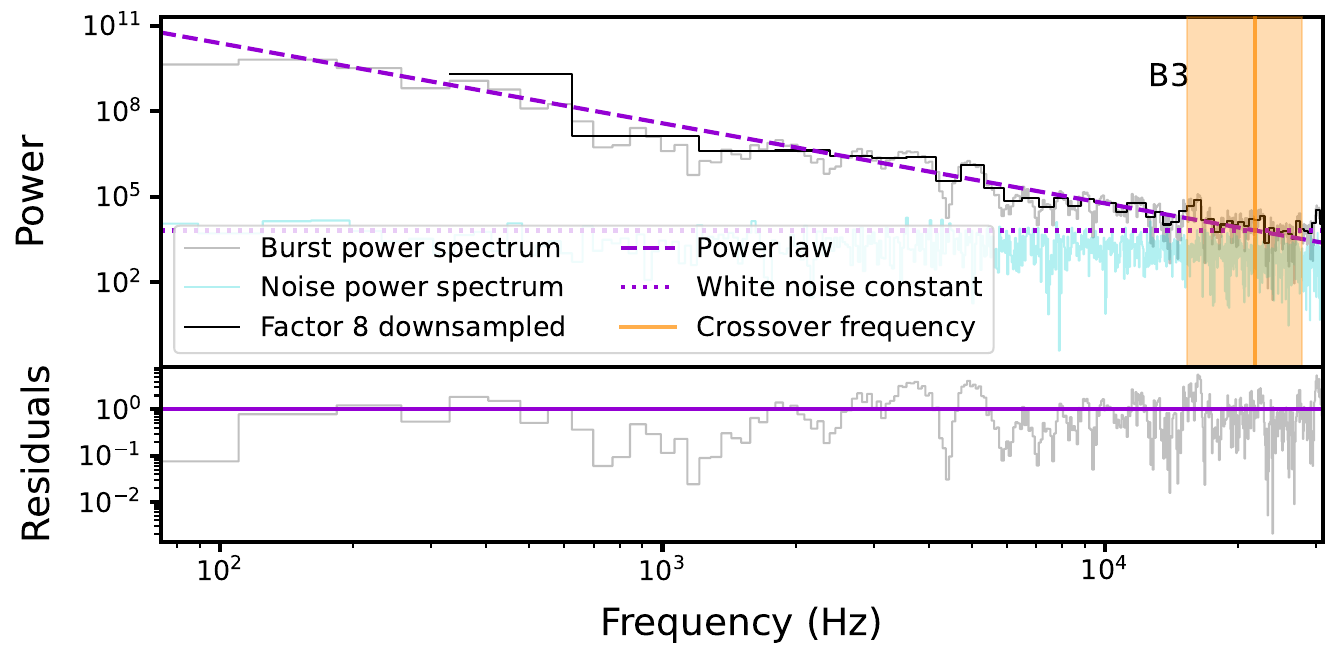}}\quad
      \subfloat[][]{\includegraphics[width=.48\textwidth]{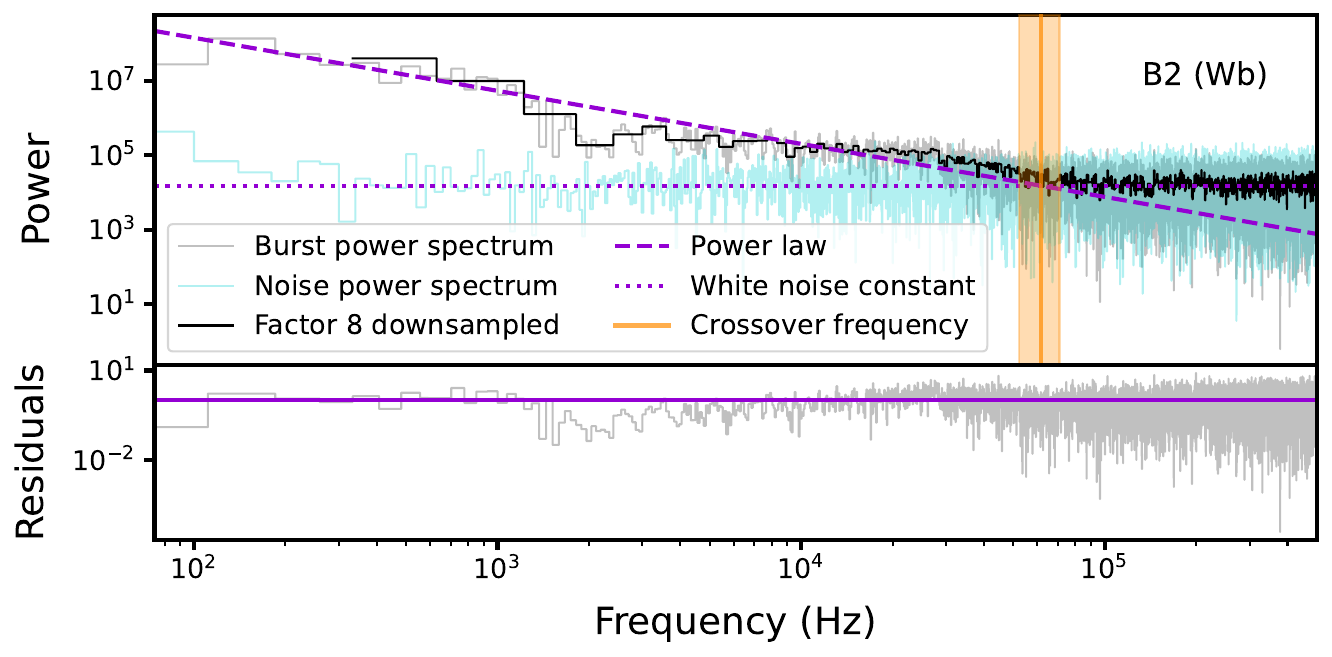}}
    \caption{The power spectra of the NRT-detected bursts from \frb\ are shown in grey, the black line shows the same power spectra but downsampled by a factor of 8. The power spectrum of the off-burst noise is shown in turquoise. No normalization has been applied to any of the power spectra. Burst IDs are labeled in the top right of each sub-figure. For B2, we also show the spectra using the simultaneous Westerbork (Wb) voltage data, which provided 1\,$\mu$s time resolution for this analysis. The burst power spectra have been fit with a model which consists of a power-law (purple dashed line) and, where possible, a white noise component (purple dotted line) using the \texttt{Stingray} modelling interface \citep{huppenkothen_2019_apj}. The residuals from these fits are shown in the bottom panels of each sub-figure. The crossover frequency where the power-law and white noise components intersect are marked with a solid vertical orange line, with shaded regions to illustrate the 3-$\sigma$ uncertainty (see the main text for more detail). Note that the x-axes are different between bursts.}
    \label{fig:ps}
\end{figure*}

Here, $\omega$ is the frequency, $A$ is the amplitude, $\alpha$ is the power law index and $C$ is a white noise constant. We assume flat priors for $A$ and $\alpha$, and a normal prior for $C$. In the case of B1 and B2 with the NRT data, $C$ could not be well constrained, so we only fit a power law without a white noise constant. These power law fits and residuals are shown in Figure~\ref{fig:ps}. The power law indices we estimate are presented in Table~\ref{tab:more_properties}. These values are comparable to those of \rthree\ and \rmeightyone\ \citep{nimmo_2021_natas,nimmo_2022_natas}. The power laws for B1 and B2 are flatter than that of B3, which is in agreement with B1 and B2 having more significant power at higher frequencies (shorter timescales) due to the presence of microshots. Furthermore, these fits allow us to investigate the highest fluctuation frequencies and to determine if there are any significant outliers from the power-law fit that may suggest the presence of QPOs. There are, however, no such outliers and we find no evidence for any QPOs in any of these three bursts.

In an effort to quantify the shortest timescales present in the burst data, we determine the crossover frequency $\omega_{c}$ where $A\omega^\alpha$ and $C$ intersect, i.e. the frequency where the amplitude of the source variations is smaller than the amplitude of the instrument noise and can no longer be measured. For B1 and B2, we were unable to constrain $C$ in the NRT data, and $\omega_c$ potentially lies at frequencies higher than we are capable of probing (i.e., timescales shorter than 16\,$\mu$s).  In addition to the power spectra and fits of the bursts in Figure~\ref{fig:ps}, we also show a power spectrum of off-burst noise in turquoise. The statistical properties of the off-burst noise are different from those of the on-burst noise due to amplitude modulation. We consequently do not normalize the spectra, and while the comparison is imperfect, it is sufficient to demonstrate that even at the highest fluctuation frequencies the burst power is well above the noise for B1 and B2. This suggests significant burst structure on timescales comparable to, and even smaller than, the time resolution of our NRT data (16$\,\mu$s), which is consistent with the presence of microshots in these bursts. For B3, which lacks any obvious microshots, we were however able to constrain $C$ and estimate $\omega_{c}$ to be $22(6)$\,kHz ($\sim$45\,$\mu$s), while for B2 with the Westerbork data (1\,$\mu$s time resolution) $\omega_{c}$ is 61(9)\,kHz ($\sim$16\,$\mu$s). These values and their uncertainties are indicated by the orange vertical lines and shaded regions in Figure~\ref{fig:ps}. One can also see by eye that this is approximately the fluctuation frequency where the on- and off-burst power spectra intersect. The microshots are thus marginally resolved by the time resolution of the NRT data, and the notches present in B3 are of the same duration as some of the microshots.

\subsubsection{High time resolution analysis}

\begin{figure*}
    \centering
    \includegraphics[width=\textwidth]{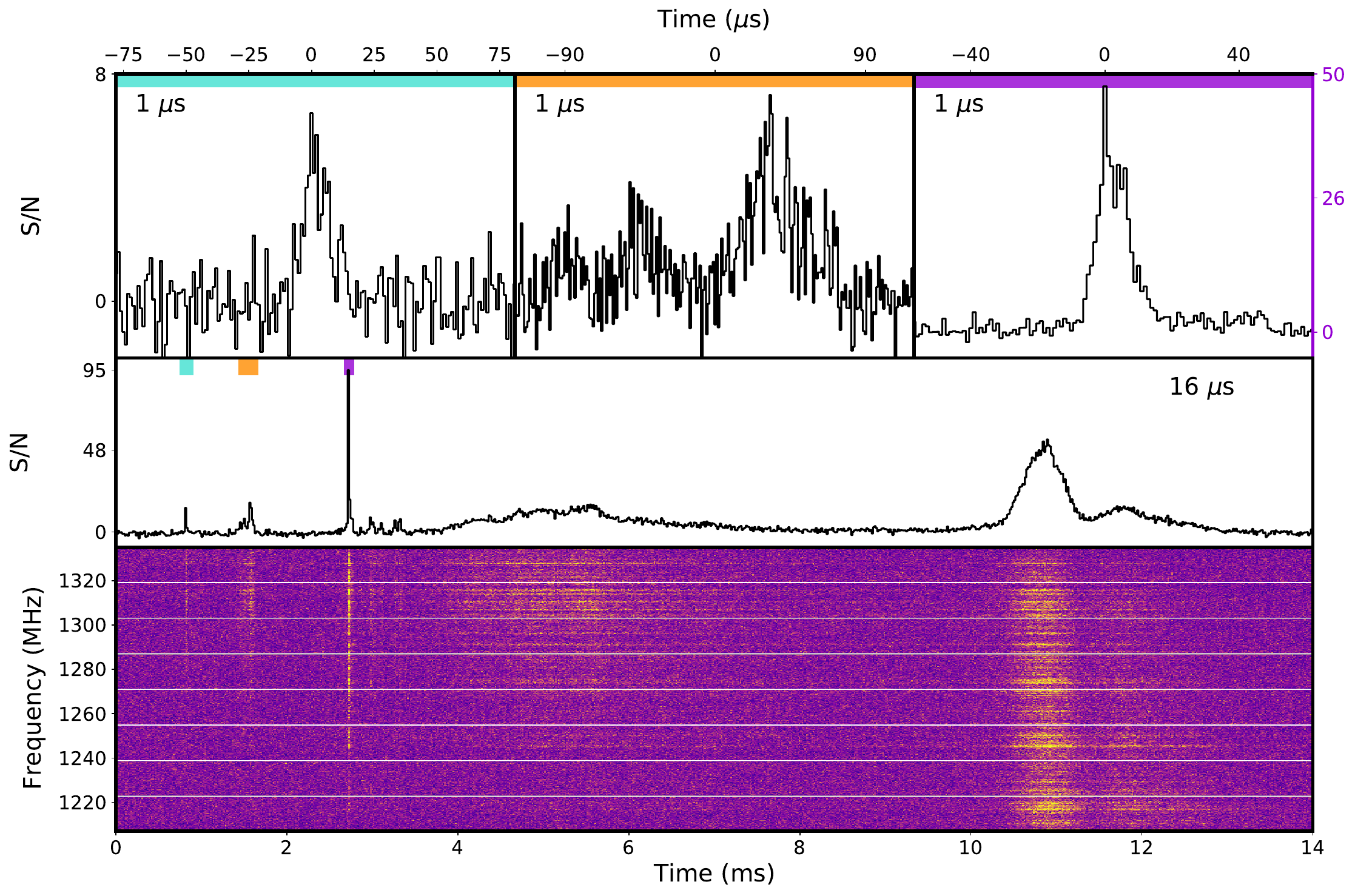}
    \caption{The bottom panel shows a zoom-in of the dynamic spectrum of burst B2, which was simultaneously detected (along with NRT) by the Westerbork RT-1 dish. The time and frequency resolution are 16\,$\mu$s and 62.5\,kHz, respectively, and the three brightest microshots are clearly visible. Above the dynamic spectrum, the frequency-averaged profile is shown. The top three panels show zoom-ins on each of these microshots (indicated by the coloured horizontal bars), at 1\,$\mu$s time resolution. These high time resolution views of the microshots confirm that we are (mostly) resolving these microshots even with the 16-$\mu$s time resolution of the NRT data. Note that the S/N of the microshots indicated by the turquoise and orange bars is shown by the left y-axis, whereas the S/N of the microshot indicated by the purple bar is shown on the right y-axis (in purple).  }
    \label{fig:wb_zoom_ds}
\end{figure*}

Using the raw voltage data recorded from the Westerbork observations, we can construct data products using \texttt{SFXC} (as mentioned earlier) with even better time resolution than the NRT data to confirm that we are marginally resolving the microshots and to determine if there is any structure on timescales shorter than the time resolution of the NRT data.  In Figure~\ref{fig:wb_zoom_ds} we show a dynamic spectrum of B2 which was detected using the Westerbork RT-1 dish. The plotted data have the same time resolution as the NRT data (16\,$\mu$s) and a frequency resolution of 62.5\,kHz. The top 3 panels show zoom-ins of the brightest three microshots at a time resolution of 1\,$\mu$s. Here it can be seen that the microshots are mostly being resolved in the NRT data, as they do not completely break up into more components. There are, however, hints of brightness fluctuations on the order of 1\,$\mu$s timescales, particularly in the third, brightest microshot (indicated by the purple bar). 

\begin{figure*}
    \centering
    \includegraphics[width=\textwidth]{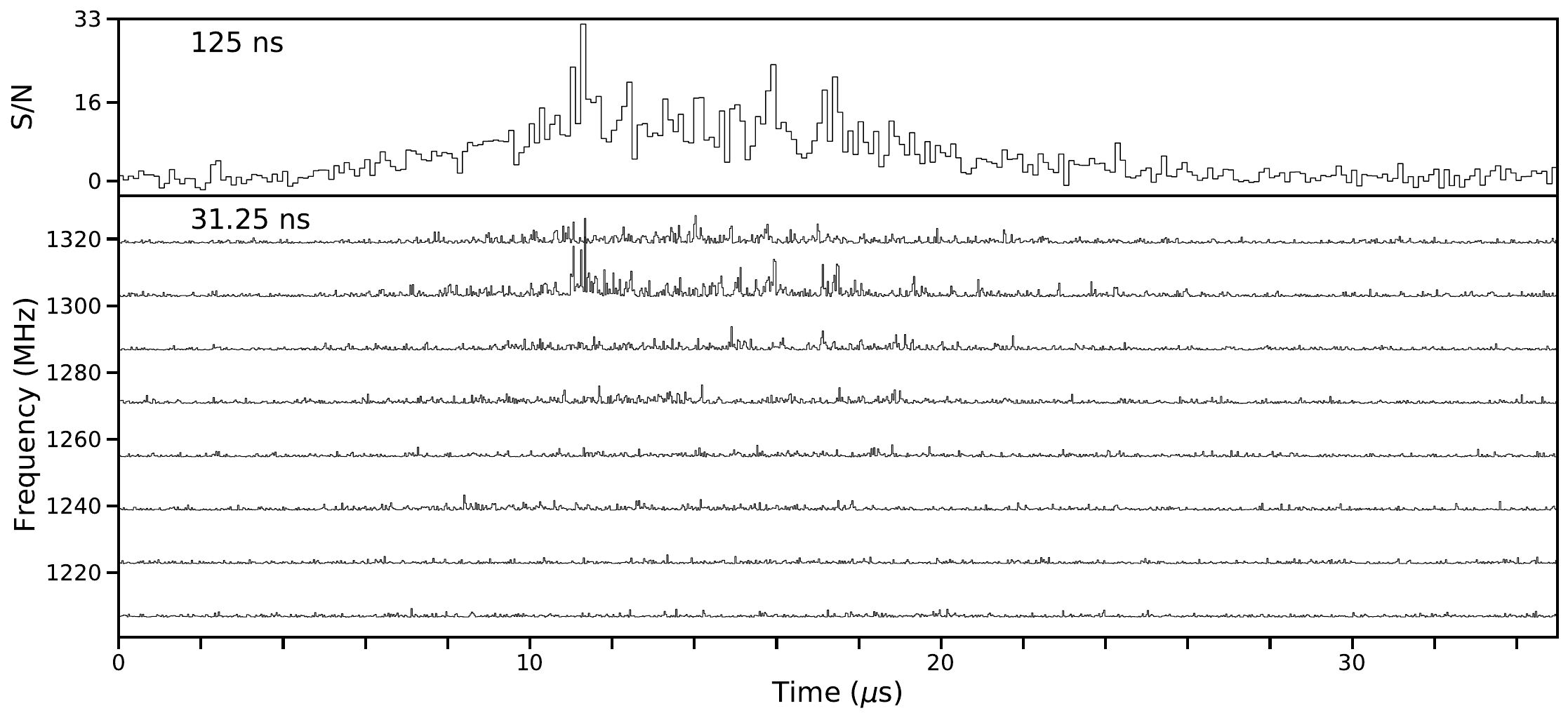}
    \caption{The top panel shows the brightest microshot in burst B2 at 125\,ns time resolution. In the bottom panel the time profile for each of the eight subbands is shown at a time resolution of 31.25\,ns, which is the best possible time resolution we can obtain with the Westerbork data. The majority of the emission is concentrated towards the top of the Westerbork observing band.}
    \label{fig:wb_zoom_jd}
\end{figure*}

In Figure~\ref{fig:wb_zoom_jd}  we further zoom in on this specific bright microshot. Instead of plotting a conventional dynamic spectrum (with poor frequency resolution), we opt to show the time series per subband sampled at the Nyquist rate at a time resolution of 31.25\,ns and a frequency resolution of 16 MHz. We find that the emission of the burst is mostly concentrated in the form of bright scintils in the top subbands. The top panel shows the time series of the microshot at a time-averaged resolution of 125\,ns. Here we can see hints of sub-microsecond structure, but due to S/N limitations, the precision to which we know the true DM and the inferred scattering timescale of about 0.5\,$\mu$s, we are unable to make concrete claims. 

\subsection{Clustering of microshots}
\label{sec:clustering}

\begin{figure*}
    \centering
    \subfloat[][]{\includegraphics[width=.475\textwidth]{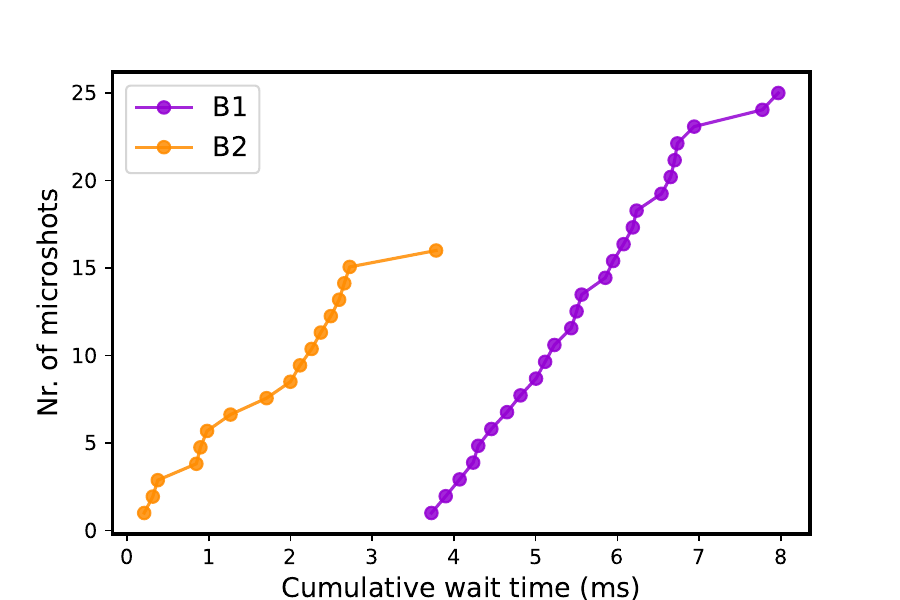}}\quad
    \subfloat[][]{\includegraphics[width=.475\textwidth]{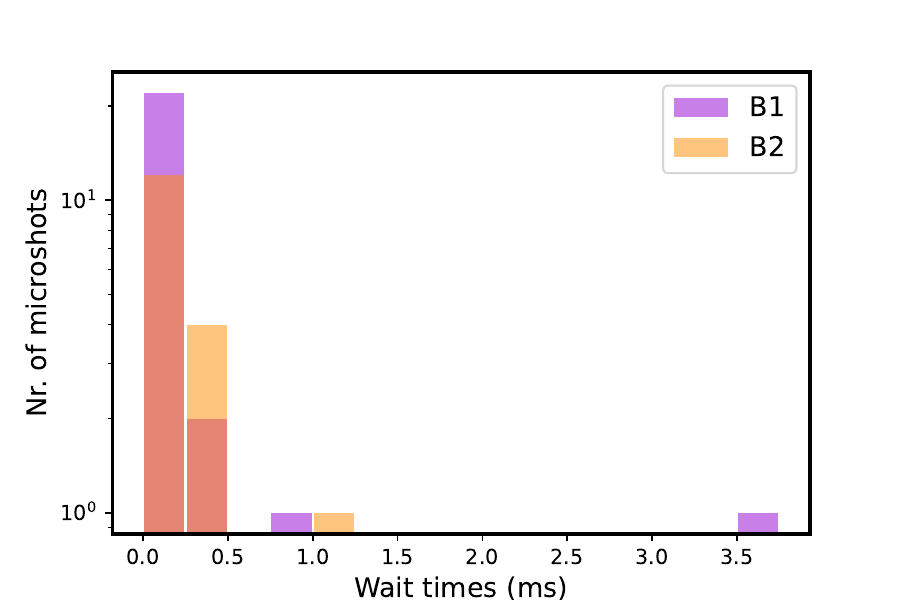}}
    \caption{\textit{(a)} The cumulative microshot distribution as a function of time since the start of the burst, for bursts B1 and B2. \textit{(b)} The wait-time distribution of the microshots identified in B1 and B2. Each histogram bin is 0.25\,ms wide.}
    \label{fig:microshot_waittimes}
\end{figure*}

To explore whether the microshots are clustered in time during the bursts, or exhibit a random (Poisson) distribution, we manually identified 27 microshots in B1 and 18 microshots in B2, and calculated the wait-time distributions of these microshots (see Figure~\ref{fig:microshot_waittimes}). The intervals $\delta$ between events of a Poisson process with a constant rate $r$ follow an exponential distribution:

 \begin{equation}
     f(\delta | r) =  r e^{-\delta r}
 \end{equation}

In the context of FRBs, \cite{oppermann_2018_mnras} have discussed how this can be generalized as a Weibull distribution:
\begin{equation}
    f(\delta|k,r)=\frac{k}{\delta}(\delta r\Gamma(1+k^{-1}))^ke^{-(\delta r\Gamma(1+k^{-1}))^k}
\end{equation}
where $k$ is the shape parameter, $r$ is the rate parameter, and $\Gamma$ is the gamma function. If $k=1$, the equation reduces to the exponential form and the events are Poisson distributed. The smaller the value of $k$, the more the distribution deviates from a Poisson distribution and the more clustered the events are.

Here we test whether the wait times between the microshots identified in these bursts are consistent with a Poissonian distribution (i.e., the posterior distribution is consistent witk $k=1$) by following the methodology of \cite{oppermann_2018_mnras} \citep[as also implemented in][]{kirsten_2021_natas,good_2023_apj,nimmo_2023_mnras}. Using a uniform prior distribution $f(k,r)$, we calculate the posterior distribution as:
 
\begin{equation}
    \mathrm{Post}(k,r|d)\propto L(d|k,r) f(k,r)  
\end{equation}

\begin{figure*}
    \centering
    \subfloat[][]{\includegraphics[width=.475\textwidth]{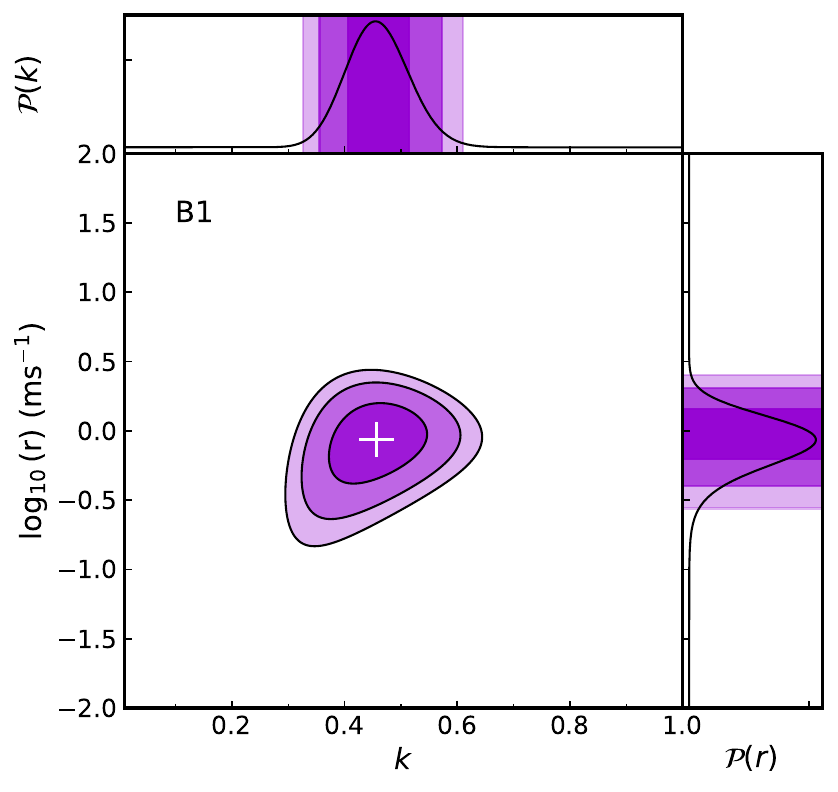}}\quad
    \subfloat[][]{\includegraphics[width=.475\textwidth]{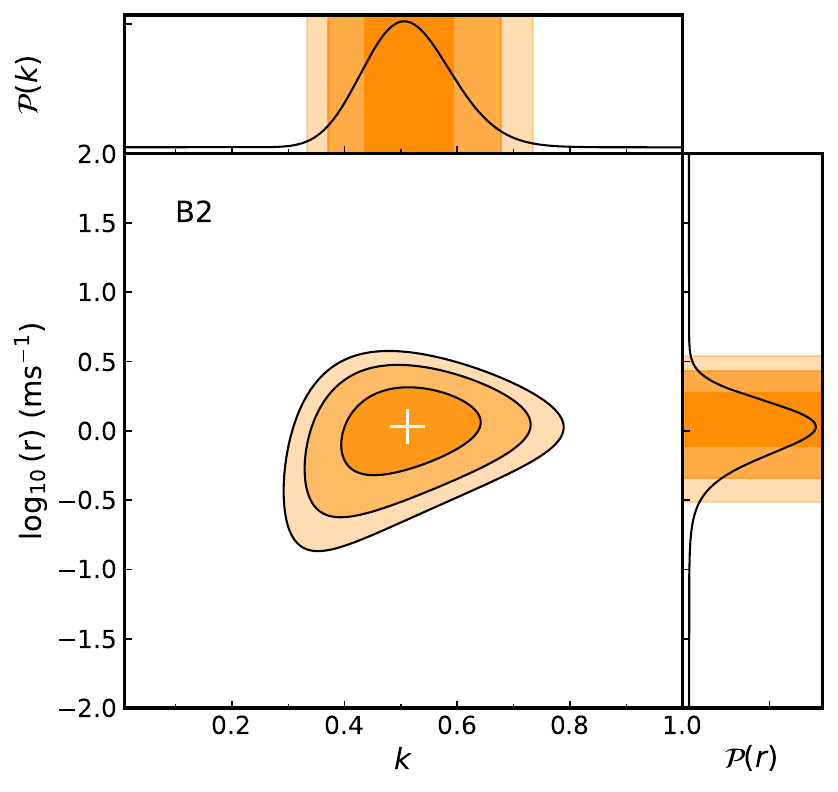}}
    \caption{Posterior probability distributions of the rate ($r$) and shape parameters ($k$) for the Weibull distributions of the microshots identified in B1 ({\it a}) and B2 ({\it b}). The contours and shaded regions show the 68\%, 95\% and 99\% confidence intervals. The white plus indicates the position of the maximum probability density.}
    \label{fig:post}
\end{figure*}

\begin{table*}

\caption{Burst properties obtained from the power spectrum, clustering, and polarimetric analyses.}
\begin{center}
{\begin{tabular}{lrrrrrr} \hline
Burst ID & $\alpha^a$ & $r$  & $k ^b$  & RM$_\text{obs}$ & $\delta$RM$_\text{obs}$    & Circular polarisation \\
 & &(ms$^{-1}$)$ ^b$ &  & (rad\,m$^{-2}$)$ ^c$ & (rad\,m$^{-2}$)$ ^d$   & fraction \\
\hline
B1  &   2.08(8)      &    $0.87^{+1.64}_{-0.59}$     &    $0.46^{+0.15}_{-0.13}$     &  $-0.93$(5) & 0.67(5)  & 4(1)  \\
B2   &   1.5(2)     &    $1.07^{+2.40}_{-0.76}$     &    $0.51^{+0.22}_{-0.18}$    &  1.10(7)    &  0.73(5)   & 6(5) \\
B3    &   2.8(2)    &    -     &    -    &  0.12(7)    & 0.75(3)    & 7(2) \\
\hline
\label{tab:more_properties}
\end{tabular}}\end{center}
$^a$ Power-law index determined by fitting a power-law model to B1-B3 using the NRT data (see the main text for more detail).
\\
$^b$ These uncertainties are the 99 \% confidence intervals.
\\
$^c$ These are the FWHM/SNR uncertainties reported by \texttt{rmfit}.
\\
$^d$ The amount of ionospheric Faraday rotation estimated using \texttt{IonFR} \citep{sotomayorbeltran_2013_ascl}.
\\

\end{table*}

 where $L(d|k,r)$ is the likelihood of the data $d$. The posterior probability distributions of $r$ and $k$ for the microshots in B1 and B2 are shown in Figure~\ref{fig:post} and the values are presented in Table~\ref{tab:more_properties}. For both bursts, $k<1$, thus providing evidence that the microshots are clustered within the bursts (and possibly even to a comparable degree in both bursts).

\subsection{Polarimetry}

We calibrated the polarimetric data by using \texttt{PSRCHIVE} tools\footnote{\url{psrchive.sourceforge.net}} \citep{hotan_2004_pasa}.  Archive files containing full polarisation information were created for the bursts using \texttt{DSPSR} \citep{vanstraten_2011_pasa}. These archives were then calibrated using the \texttt{pac} command from \texttt{PSRCHIVE} in combination with a 3-Hz pulsed noise diode scan and a polarisation calibration modeling (pcm) file. The pcm files are constructed by analyzing observations of the pulsar PSR~J0742$-$2822 during which the receiver horn is rotated by $\sim$180$^{\circ}$ over the course of a 1-hr observation. Notably, we do not simply apply single-axis calibration to correct the differential gain and differential phase of the receptors, but make use of the complete reception model \citep{vanstraten_2006_apj,vanstraten_2013_apjs}, which in addition to the single-axis calibration, maps the frequency-dependent gain, and takes into account the non-orthogonality of the receptors to more accurately estimate the complete instrumental response (Guillemot et al., \textit{submitted}). 

\begin{figure}
    \centering
    \includegraphics[width=0.5\textwidth]{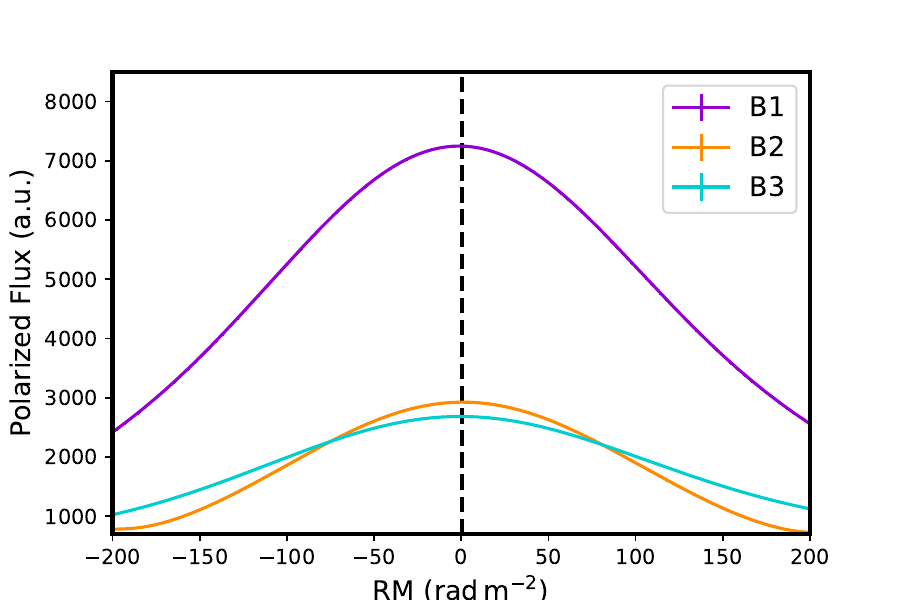}    
    \caption{The RM spectra produced by \texttt{rmfit} for the three \frb\ bursts detected with the NRT. The optimal values for RM$_{\text{obs}}$ are $-0.93$(5), 1.10(7) and 0.12(7)\,rad\,m$^{-2}$ for B1, B2 and B3, respectively. The vertical dashed line is at 0.6\,rad\,m$^{-2}$, the RM reported for a bright burst from \frb\ by CHIME/FRB \citep{mckinven_2022_atel}.}     
    \label{fig:rfmit}
\end{figure}

To determine the RMs of the bursts, we use \texttt{PSRCHIVE}'s \texttt{rmfit}. In Figure~\ref{fig:rfmit} the obtained Faraday spectra are shown for the three bursts detected with NRT. The optimal RMs, with the FWHM/SNR uncertainties reported by \texttt{rmfit}, are all close to 0\,rad\,m$^{-2}$.
Using \texttt{IonFR} \citep{sotomayorbeltran_2013_ascl}, we estimated the amount of ionospheric Faraday rotation. All the aforementioned RM values are shown in Table~\ref{tab:more_properties}. 
Importantly, the values obtained from \texttt{rmfit} are the observed RM values (RM$_\text{obs}$), since they have not been corrected for the effect of the ionosphere or redshift.  These RM$_\text{obs}$ values are consistent with what has been reported in other works \citep{mckinven_2022_atel,feng_2023_arxiv,zhang_2023_arxiv}. The RM also appears to be relatively stable, as \cite{zhang_2023_arxiv} demonstrated that the RM changed very little over the course of 2 months. We too find no major change in the RM between B1 and B3, which are separated by 18 days.

We validated our polarimetric calibration using an observation of the pulsar PSR~B0355+54. The resulting polarimetric profiles matched those in the European Pulsar Network\footnote{\url{www.epta.eu.org/epndb/}} (EPN) database centered at 1408, 1642 and 1710\,MHz, and the RM we retrieved is $82.9(9)$\,rad\,m$^{-2}$, which when taking into account possible ionospheric variations is consistent with $81.5(3)$\,rad\,m$^{-2}$ reported in the ATNF catalogue  \citep{manchester_2005_aj}\footnote{\url{www.atnf.csiro.au/research/pulsar/psrcat/}}. 

\begin{figure*}
    \centering
    \includegraphics[width=\textwidth]{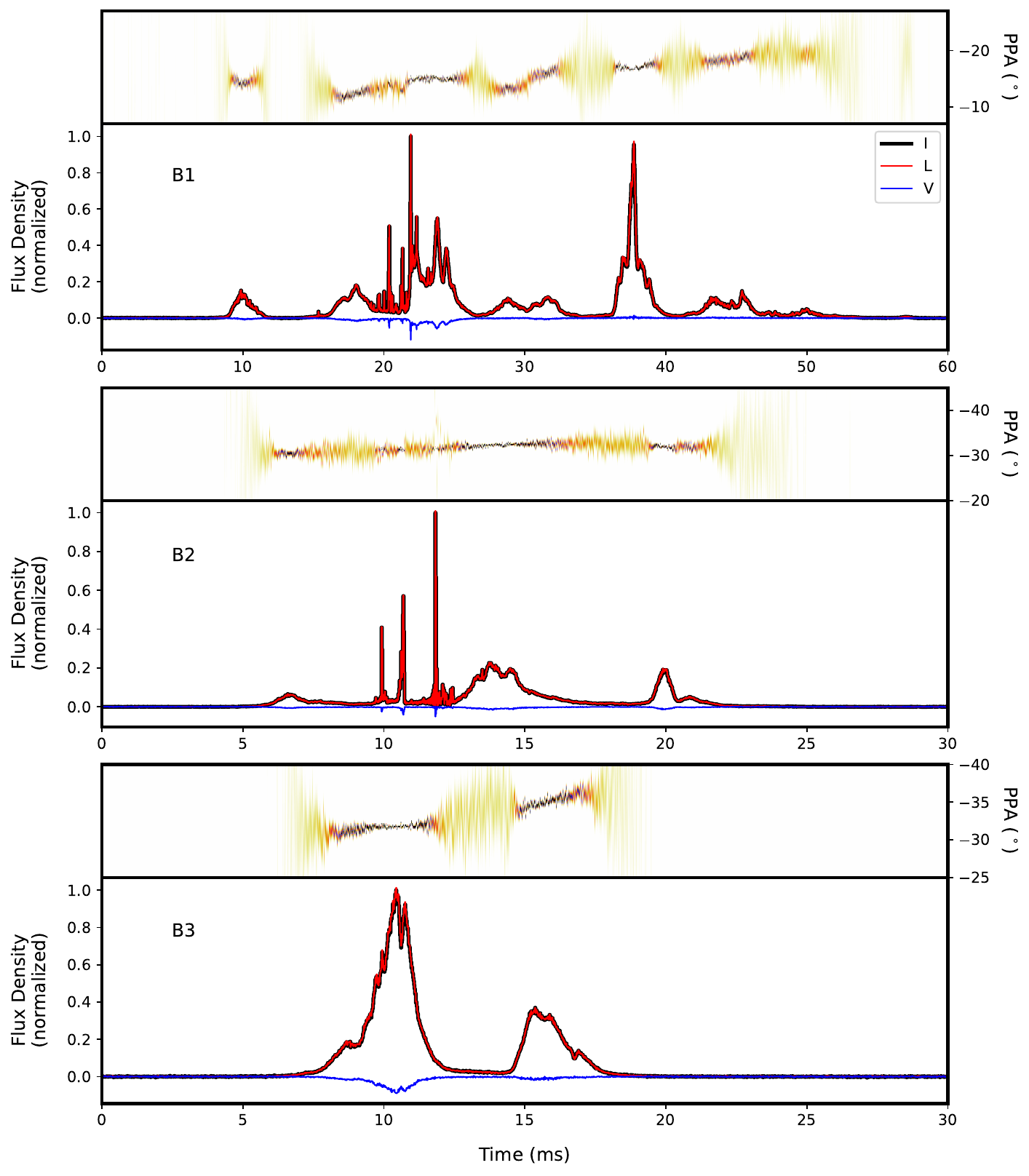}
    \caption{Polarimetric profiles of the three \frb\ bursts detected with the NRT (see burst ID in each sub-figure). The total intensity is shown by the black burst profiles and the unbiased linear and circular polarisation in red and blue, respectively. These profiles are shown at the full 16-$\mu$s time resolution of the data, and channels contaminated by RFI were flagged before averaging over frequency to obtain these profiles. The probability distribution \citep[following ][]{everett_2001_apj} of the PPAs are shown in the panels above the burst profiles, and are zoomed-in to show subtle variations with time. These profiles have been polarimetrically calibrated using the full receptor model and de-Faradayed with an RM of 0.6\,rad\,m$^{-2}$. }
    \label{fig:polarimetry}
\end{figure*}

Figure~\ref{fig:polarimetry} shows the polarimetric profiles of the NRT detected bursts, assuming an RM of 0.6\,rad\,m$^{-2}$. Before creating the frequency-averaged profiles, we masked the frequency channels contaminated by RFI in the burst data, as well those channels that had to be masked due to RFI in the noise diode scans. Since the polarimetric profiles and PPAs do not significantly change if we correct for the small range of RM values that we measure, we adopt this single value which was previously reported and is consistent with our results. We used unbiased linear polarisation \citep{everett_2001_apj} for our analyses and plots,
\begin{equation}
    \text{L}_{\text{unbiased}}=\Bigg\{
    \begin{array}{lr}
    \sigma_I \sqrt{(\frac{\text{L}}{\sigma_I})^2-1} & \rm{if} \quad \frac{L}{\sigma_I} \geq 1.57\\
    0 & \rm{otherwise}
    \end{array}
\end{equation}

The bursts are all about 100\% linearly polarised with hints of circular polarisation in some components. The circular polarisation fractions are tabulated in Table~\ref{tab:more_properties}. We do not believe these inklings of circular polarisation to be astrophysical, since the circular polarisation fraction of the bursts still retains a slight frequency dependence despite our careful calibration procedure (see Figure~\ref{fig:stokes_spectra}). Nevertheless, this imperfection in the calibration has no significant effects on our results or interpretation thereof. Unlike many other repeating FRBs which show PPAs that are flat across, and even between, bursts \citep[e.g.,][]{nimmo_2021_natas}, the PPAs of these \frb\ bursts show some variability. The PPAs in B1 and B3 display swings and vary by approximately 10 degrees across the burst duration. We also see significant jumps in the PPA at the time of the dense microshot forest.  \cite{feng_2023_arxiv}  also reported PPA swings in some bursts from \frb, akin to those seen in FRB~20180301 \citep{luo_2020_natur}.

\section{Discussion} \label{sec:discussion}

\subsection{On dedispersion}
\label{subsec:dm}

The DM is a fundamental quantity associated with FRB sources, and knowledge of the correct DM is critical for measuring many other attributes such as burst widths, peak flux densities and scattering timescales, to only name a few examples. Determining the DM accurately, however, is non-trivial as has been noted by various other authors  \citep[e.g.,][]{hessels_2019_apjl}. 

One can opt to determine the DM by trying to maximise the frequency-averaged S/N of a burst, as is effectively done in most burst searches. This method unfortunately often results in overestimation of the DM due to downward drifting complex frequency-time structure \citep[`sad-tromboning';][]{hessels_2019_apjl} which is present in many FRBs (especially repeaters) and is most likely not related to dispersion caused by intervening media. \cite{hessels_2019_apjl} followed an alternate approach which involves trying to optimise for burst structure by maximizing the time derivative of the peaks in frequency-averaged burst profile. This method has been widely adopted through its implementation in the \texttt{DM\_phase} package \citep{seymour_2019_ascl} and other similar techniques. While this greatly improves upon simple S/N-maximization, it might still fail to determine an accurate DM for weaker bursts or bursts which lack any prominent structure.

Fortunately, bursts B1 and B2 exhibit exquisite structure. There are exceptionally bright components present that are at most only a few bins wide in the NRT data (tens of microseconds in duration). As was mentioned in Section~\ref{sec:analysis}, we determined the DM of the bursts presented in this paper by aligning the microshots so that the emission occurs simultaneously at all frequencies. The DM estimates obtained from using \texttt{DM\_phase}, with some manual guidance on what range of times and fluctuation frequencies to use, are consistent with our DM values. 

Importantly, in accurately determining this DM a few assumptions are made that require justification. Firstly, we are assuming that the entire burst, and all of its components, have a single DM value. Furthermore, the structure optimization approach assumes that in the reference frame of the source, burst emission occurs simultaneously across a broad range of frequencies. Building upon these two assumptions, we further assume that it only holds true for the microshots, but not necessarily for broader components/sub-bursts. This is motivated by our observation of a residual drift that can be seen in the broader sub-bursts after correcting for dispersion (using a DM which we measure using the microshots). While, effectively, a single DM value temporally aligns the emission of multiple microshots across all observed frequencies, it does not vertically straighten the broader components in the dynamic spectrum. We have overdedispered bursts B1 and B2 to straighten these broader burst components, ignoring the microshots, and find that no single DM value can perfectly straighten all of the broader burst components, as they have slightly different (apparent) DMs. These DMs are always slightly larger than the DM determined using the microshots, but smaller than about $0.3$\,pc\,cm$^{-3}$. Other authors have also noted that for some bursts not all components can be temporally aligned using a single DM \citep[e.g.,][]{michilli_2018_natur,zhou_2022_raa}.  In addition to the $\lambda^2$ dependent delay introduced by dispersion, there appears to be an additional drift which determines the spectro-temporal morphology of the sub-bursts, perhaps related to the emission mechanism. \cite{jahns_2023_mnras} and \cite{chamma_2021_mnras} have noted that this drift increases linearly with burst duration, and that the sad-trombone effect occurs within sub-bursts, as well as between them.

Alternatively, if we refrain from making our initial assumption and allow for different burst components to have different DM values, the residual drift seen in broader burst components after dedispersion (to the microshot-determined DM) could perhaps be explained by a difference in column depth or electron density between the microshots and the broader burst components. Considering a scenario where FRBs are emitted within an ionised medium around a compact progenitor, the difference in DM would arise from the broader components being emitted deeper within the ionised medium closer to the progenitor, and thus travelling along a longer path or through denser media than the microshots. Furthermore, within close proximity to a magnetar progenitor, for example, the plasma becomes relativistic and nonlinear propagation effects start to play a significant role. The most important nonlinear propagation effect in magnetar winds is the filamentation instability \citep{sobacchi_2022_mnras}, which is expected to result in slight variations of the effective DM, on the order of the variations we see in different millisecond-duration components in these bursts.

Now, considering a scenario where the position of the burst emission can be explained by radius-to-frequency mapping \citep{lyutikov_2020_apj,tong_2022_mnras}, it is more natural to assume that the burst emission of some components is delayed at lower frequencies as a result of material that is propagating radially outward from a central engine, than assuming the inverse situation where emission is delayed at higher frequencies due to the emission region moving towards the central engine. Consequently, if a burst is found to have a DM that results in dedispersed emission occurring later at higher frequencies, this DM is likely overestimated.

In short, we thus argue that the correct DM of a burst, is one that temporally aligns short timescale emission at all observed frequencies, but may still cause longer duration sub-bursts to have a drift where emission is delayed at lower, but not higher, frequencies. The shortest timescales in a burst should thus be used to quantify the uncertainty of the DM measurement. Methods that enable accurate quantification of the uncertainty on the structure metric of a burst, such as \cite{sutinjo_2023_arxiv}, are especially useful in this regard.

Reporting accurate uncertainties is especially important when individual DM measurements are made for a large numbers of bursts.  Negligence in this regard may result in nonphysical interpretations, as significant changes in the DM of bursts within a short period of time requires quite extreme variations in the line-of-sight plasma local to the source (e.g., very dense filaments moving across the line-of-sight). In the case of the Crab pulsar, DM variations of  $\sim$0.02 pc\,cm$^{-3}$\,month$^{-1}$ have been seen that can be interpreted as a scattering screen caused by filaments in the nebula \citep{driessen_2019_mnras}. While some FRBs do originate from a very extreme magneto-ionic environment, and large DM and RM variations between bursts are expected, true astrophysical DM variations of a few units over the course of a few weeks seem unlikely. We therefore argue that it is in general better and more astrophysically motivated (although still not ideal) to dedisperse multiple bursts that were detected from the same source within a period of weeks, to a DM which has been  determined from burst(s) that have short timescale structures, than it to dedisperse each burst to an individual DM (regardless of the method used to determine these inidividual DMs). 

Notably, however, there are some sources, such as \rone\ have been seen to have credible DM variations on longer timescales. The DM has been seen to increase by approximately 4\,pc\,cm$^{-3}$ over the course of $\sim$3 years \citep[][]{michilli_2018_natur,hilmarsson_2021_apjl}, before decreasing by about 10\,pc\,cm$^{-3}$ over the next 3 years \citep{wang_2022_atel_15619}.

\subsection{Microshots in \frb\ }
\label{sec:microshots}
Two of the three \frb\ bursts presented in this paper consist of burst components with durations ranging from about ten microseconds to a few milliseconds. Summing over the entire 512-MHz observing bandwidth, the brightest of these microshots in B2 (when dedispersed using a DM of 219.377\,pc\,cm$^{-3}$)  has a S/N above 2000, corresponding to a peak flux density of approximately 450\,Jy. These microshots appear to be mostly resolved at around 10\,$\mu$s, but there are hints of sub-microsecond time structure, as shown in Figure~\ref{fig:wb_zoom_jd}, that require confirmation.

In addition to the microshots there are also more typical wider burst components. Interestingly, even after the burst has been corrected for dispersion and the microshots are aligned so that emission occurs simultaneously at all frequencies, the wider burst components still exhibit a residual drift apparently not associated with dispersion (see, e.g., the components between 5 and 6\,ms in B1 of Figure~\ref{fig:microshots}). Assuming the microshots with comparable bandwidth are also drifting, on a conservative estimate of 5 time bins, this results in a drift rate which is at least a factor of 6 larger than the more typical sub-bursts.

The range of timescales we are probing is also evident in the NRT burst power spectra, where there is power significantly above the noise level for frequencies ranging from $\sim 40 - 3\times10^4$\,Hz (corresponding to timescales of $\sim 32\,\mu$s $-$ 25\,ms) for B1 and B2. For B3, which lacks the microshots, the crossover frequency starts becoming visible and suggests the shortest timescales on which there is significant burst structure is about $35-62\,\mu$s, above the duration of the shortest microshots but corresponding to the duration of the notches. The Westerbork B2 power spectrum indicates that the shortest timescale where there is significant structure is around $14-19\,\mu$s, explaining why the crossover frequency could not be determined in the NRT data, but that the shots are being marginally resolved.

Furthermore, we have shown in Section~\ref{sec:clustering} that these microshots are not homogeneously spread out through the burst, but are clustered into dense forests that can be described by a Weibull shape parameter of about 0.5. The different timescales of burst emission and different drift rates, together with the clustering of the microshots, require theoretical understanding

\subsection{In comparison with FRBs}
It is well known that temporal structure much shorter than the burst width can be present in FRBs. \cite{farah_2018_mnras} showed that FRB~170827 has structures that are only $\sim30\,\mu$s in duration, and the total duration of one of the \rone\ bursts in \cite{michilli_2018_natur} is only $\sim30\,\mu$s. \rthree\  bursts can have structure on the order of a few microseconds \citep{nimmo_2021_natas}, and even shorter timescale structure (tens of nanoseconds) is present in some \rmeightyone\ bursts \citep{nimmo_2022_natas}. Additionally, in a search of archival GBT data, \cite{snelders_2023_arxiv} detected isolated bursts from \rone\ that are only a few microseconds in duration. At least some repeaters are thus capable of producing such short timescale emission independent from broad-timescale emission. We have shown that the \frb\ bursts presented in this work have structure on timescales down to a few microseconds, and potentially even lower (see Figure~\ref{fig:wb_zoom_jd}), adding \frb\ to the handful of FRBs that exhibit known microstructure. 

However, we do not see these forests of microshots in all bursts from \frb. Similarly, \cite{nimmo_2021_natas,nimmo_2022_natas,nimmo_2023_mnras} noted that only some of the bursts have microstructure. This raises the question if these microshots are in fact a type of burst which can sometimes occur alone as with \rone\ or FRB~170827, and at other times simultaneously with a different broader type of burst, as with B1 and B2 presented here. While the isolated microsecond duration FRBs might be `tips of an iceberg' --- extraordinarily bright microshots (much brighter than the ones shown in this work) from a broader burst too faint to be detected --- this seems unlikely because of the fine-tuning that is needed to see microshots but no broader components \citep{snelders_2023_arxiv}. Regardless, investigating emission on a wide range of timescales provides valuable constraints for theoretical models. 

The occurrence of microshots is not easily explained by propagation effects such as fragmentation instabilities \citep[e.g.,][]{lu_2022_mnras}. One of the main arguments for this is the lack of explanation as to why the instability would cause the breakdown of the emission into the microshots, while not affecting the broader emission components that occur mere milliseconds later. Ignoring relativistic effects, these timescales correspond to small emission regions which are more naturally explained by magnetospheric magnetar models \citep{beniamini_2020_mnras_498} than the synchrotron maser models where emissions is generated in a shock far away from the central engine \citep[e.g.,][]{metzger_2019_mnras}.

\subsection{In comparison with neutron stars}
There are five known Galactic radio-emitting magnetars \citep{camilo_2006_natur,camilo_2007_apjl,levin_2010_apjl,eatough_2013_natur,karuppusamy_2010_aa}. These magnetars emit radio pulsations that while semi-stable in rotational phase, show pulse-to-pulse variability in flux density, spectral index and the shape of the pulse profile \citep[e.g.,][]{lazaridis_2008_mnras,serylak_2009_mnras}. Similarly to the bursts seen from repeating FRBs, the individual magnetar radio pulses (as well as the average pulse profile) tend to be highly linearly polarised, while having circular polarisation up to a few tens of percent \citep[e.g.,][]{kramer_2007_mnras}, although this too can be time variable \citep{dai_2019_apjl}. Furthermore, individual radio pulses from magnetars have often been described as `spiky' in the literature \citep[e.g.,][]{serylak_2009_mnras,caleb_2022_mnras}. The pulses jump around in phase within the larger integrated profile, and are typically a few milliseconds wide, and sometimes exceptionally bright \citep[e.g.,][]{yan_2015_apj,pearlman_2018_apj,suresh_2021_apj}. The pulses can also occasionally show hints of more complex structure, such as features as short as $\approx0.2$\,ms \citep{camilo_2006_natur} or two components with opposing spectral indices \citep{suresh_2021_apj}, further prompting comparison with FRBs. While the radio pulsations are the defining characteristic of radio magnetars, these pulsations are not necessarily ever-present and can disappear as in the case of XTE~J1810$-$197 \citep{camilo_2016_apj}, or reawaken after years of radio-quiescence \citep{lyne_2018_atel}. We require long-term monitoring of more repeaters to determine whether the burst storm behaviour is analogous to this.  Unlike FRBs that are sometimes extremely narrowband \citep[$\sim$10$-$30\% fractional bandwidths, e.g.,][]{gourdji_2019_apjl,kumar_2021_mnras}, the magnetar pulsations have broadband flat spectra \citep[e.g.,][]{lazaridis_2008_mnras,dai_2019_apjl}. The pulses also do not show any downward drifting or sad-tromboning, as is evident in some bursts from repeaters. Magnetar radio bursts have not been extensively studied on microsecond timescales, however, and the comparison with FRBs is thus incomplete. 

It is also worth noting that even the shortest timescale components in typical magnetar pulses are still longer in duration than the microshots in the \frb\ bursts presented here, or the microshots seen in giant pulses (GPs). GPs are extremely bright pulses from pulsars that last a few microseconds or less, and that occur within narrow phase windows. GPs have also been detected from radio magnetars \citep[e.g.,][]{caleb_2022_mnras}. GPs from the Crab (PSR~B0531+21) have been shown to occasionally consist of nanosecond-duration highly polarised shots of emission \citep{hankins_2003_natur}. The main pulse (MP) of the Crab has also been shown to resolve down to a dense forest of broadband microsecond bursts, which sometimes can be further resolved into extremely bright ($>10^3$\,Jy),  highly polarised and narrow-band nano-shots with sub-ns durations \citep{hankins_2007_apj}. While morphologically similar to the spiky emission in magnetars and the microshots in FRBs, it is important to note that the timescales are much shorter for the GPs. Furthermore, the polarimetric properties of the nanoshots in the MP can change drastically from one shot to another, meaning that with sufficient averaging or a high enough density of nanoshots, the polarization washes out and the profile becomes only weakly polarized.  As pointed out by other authors, pulses from the High-Frequency Interpulse (HFIP) share more similarity with the repeaters in that they are often highly linearly polarised ($\sim80-100\%$), have little to no circular polarisation and have polarisation positions angles that remain constant not only throughout the pulse, but also between pulses \citep{hankins_2016_apj}. This thus proves to show that magnetised neutron stars such as the Crab are capable of producing burst emission with very diverse properties.

The only Galactic radio burst brighter than GPs, are the extremely bright isolated radio bursts that have been detected from the Galactic magnetar SGR~1935+2154 \citep{bochenek_2020_natur,chime_2020_natur_galacticfrb}. The brightest of these bursts had an equivalent isotropic energy approximately a thousand times brighter than any previously detected radio magnetar burst; this burst was indeed even more luminous than bursts detected from the extragalactic \rmeightyone\  \citep{nimmo_2022_natas} situated approximately 3.6\,Mpc away \citep{bhardwaj_2021_apjl,kirsten_2022_natur}. It was further shown in follow-up observations that SGR~1935+2154 is also capable of producing fainter bursts, spanning 7 orders of magnitude in energies with comparable rates \citep{kirsten_2021_natas}. All together, this shows that at least some fraction of the FRB population is produced by magnetars. Assuming a magnetar progenitor, the discovery of an FRB source in a globular cluster \citep{kirsten_2022_natur}, does however beckon for formation channels other than core-collapse supernovae, such as accretion-induced collapse or binary merger.

\subsection{In comparison with solar radio bursts}
The Sun produces various types of solar radio bursts (SRBs)  that bear striking morphological similarities with FRBs, albeit on much longer timescales and with much lower luminosity \citep[for a review see, e.g.,][]{wild_1963_araa,pick_2008_aarv}. In fact, \cite{lyutikov_2002_apjl} predicted coherent radio emission similar to SRBs during X-ray bursts from soft gamma-ray repeaters, by virtue of comparison with solar flares. The first three types of SRBs were coined by \cite{wild_1950_ausra}. Type I SRBs (also called `noise storms') can last from hours to days, with the emission typically observed between $50-500$\,MHz and are made up of short-duration ($\sim1\,$s) and narrow bandwidth bursts which occur on top of continuum emission. Type II and Type III SRBs have more complex structure and exhibit time-frequency drifting on the order of $\sim0.25$ and $\sim20$\,MHz\,s$^{-1}$, respectively. Type II SRBs last approximately minutes and are associated with coronal mass ejections that create a shock that propagates outward through the solar corona, while Type III SRBs last only tens of seconds and are associated with high-velocity electrons that are accelerated along open magnetic field lines resulting from magnetic reconnection events that cause solar flares. Ignoring the duration of the emission, the downward drifting structure of Type II and III bursts looks very similar to repeater FRB emission, and also specifically to the broader emission components and microshots seen in the \frb\ bursts presented here. However, it is worth noting that SRBs can be  significantly circularly polarized \citep[e.g.,][]{kansabanik_2022_soph}, whereas repeaters are more typically highly linearly polarized. This could potentially be attributed to the emission being relativistic in the latter case. There are also other types of SRBs and more elaborate sub-classifications, but discussion thereof is not required for the comparison being made in this work.  

The different types of SRBs are produced by different emission mechanisms, and while the emission physics are naturally vastly different from that of FRBs which are many order of magnitudes more energetic and shorter in duration, this provides an interesting analogy for the \frb\ bursts presented in this paper. Just as the Sun produces different types of bursts through different processes, sometimes simultaneously, FRB sources could potentially be generating different types of bursts (in terms of energy, duration, bandwidth or drift rate), that are occasionally superimposed. In the solar scenario the outward propagating shock of a coronal mass ejection results in longer duration, lower drift rate Type II bursts, but around the same time short-duration, higher-drift-rate Type III bursts can be produced along the open field lines. One of the models that has been proposed to explain magnetar X-ray bursts, and that has been adapted to also explain FRBs, is the magnetar fireball model. In this model energy arising from starquakes or magnetic reconnection is injected into a plasmoid which is enclosed by magnetic field lines close to  the surface of the magnetar. As the fireball expands along the magnetic field lines, the energy can dissipate as coherent radio emission resulting in FRBs \citep{ioka_2020_apjl_904,wada_2023_mnras}. The observed spectrum of the emission is dependent on the configuration of the magnetic field. In analogy to the SRBs, the microshots and broader drifting components we observe in these \frb\ bursts could potentially be emission with different drift rates arising from different magnetic field configurations.

\section{Conclusions \& future work}
\label{sec:conclusions}
We have observed the highly active repeating \frb\ using the Nan\c{c}ay and Westerbork radio telescopes. We analyze a small subset of bursts with extremely high S/N, dynamic range, and time resolution. These data allow us to investigate effects that may likely be present in the bursts from other FRBs, but unapparent due to their faintness or lack of time resolution used in the data recording. We come to the following main conclusions and suggestions for future work:

\begin{itemize}
    \item As previously pointed out by other authors \citep{hessels_2019_apjl}, the determination of FRB DMs is non-trivial because of frequency-dependent burst structure and time-frequency drifts that are unlikely to be related to dispersion in the intervening media. Here we have used broadband microshots in \frb\ bursts to determine the optimal DM. We find that while the microshots are well fit by a single DM, there is a residual drift in the broader burst components. This means that DM determination via frequency-averaged burst structure maximization will likely fail for bursts with lower S/N. We thus caution that DM variation in repeating FRBs is difficult to establish robustly, and that DM uncertainties should consider the shortest-observable time structures \citep{sutinjo_2023_arxiv}.
    \item After accurate correction for dispersion, we still observe residual time-frequency drifts within broader sub-bursts of \frb. This residual time-frequency drift may be related to an outwards propagating emission region and radius-to-frequency mapping \citep{hessels_2019_apjl}. Alternatively, it could be due to an additional electron column density between the emission sites of the broad and narrow (microshots) burst components; the DM required to straighten the broad burst components is $\lesssim 0.3$\,pc\,cm$^{-3}$.
    \item We show that some \frb\ bursts have brightness variations spanning from a few milliseconds to less than a microsecond. Such a broad range of timescales, within a single burst event, has also been observed in \rthree\ \citep{nimmo_2021_natas} and \rmeightyone\ \citep{nimmo_2022_natas}. Furthermore, we find that the brightest microshots in our observed bursts are clustered in time compared to the total duration of the bursts. The range of timescales and clustering of microshots requires theoretical understanding. Since the microshots show extremely high (or possibly no) time-frequency drift, it is possible that they arise from a distinct physical mechanism compared to the broader sub-bursts, which do drift.
    \item Though solar radio bursts (SRBs) are obviously very different in their luminosity and timescale, the existence of multiple types of SRBs, which sometimes occur superimposed on each other provides an interesting analogy to consider. The broadband microshots from \frb\ could be generated in a physically distinct way compared to the wider frequency-swept emission. Supporting this hypothesis, \citet{snelders_2023_arxiv} find isolated microshots from \rone, which occur at different times compared to wider, time-frequency swept bursts. However, the similar polarimetric properties between the microshots and wider sub-bursts suggests a physical connection.
    \item The timescales of the \frb\ microshots are compatible with those of giant pulses from young and millisecond rotation-powered pulsars \citep[e.g.,][]{hankins_2016_apj}, but the overall \frb\ burst durations of tens of milliseconds are much longer. While the overall durations and spiky burst structure are consistent with pulses from radio magnetars, sub-millisecond temporal structures have not been extensively studied in magnetars. We therefore encourage microsecond-resolution studies of bright radio bursts from magnetars.
\end{itemize}

\section*{Acknowledgements}

DMH would like to thank Pawan Kumar and Emanuele Sobacchi for helpful discussions about the interpretation of the results presented in this work. Research by the AstroFlash group at University of Amsterdam, ASTRON and JIVE is supported in part by an NWO Vici grant (PI Hessels; VI.C.192.045). The Nan\c{c}ay Radio Observatory is operated by the Paris Observatory, associated with the French Centre National de la Recherche Scientifique (CNRS). We acknowledge financial support from the ``Programme National de Cosmologie et Galaxies'' (PNCG) and ``Programme National Hautes Energies'' (PNHE) of CNRS/INSU, France. The Westerbork RT-1 telescope is maintained and operated by ASTRON.  We express our gratitude to the operators R. Blaauw, J.J Sluman and H. Mulder for scheduling observations. We would like to express our gratitude to Aard Keipema and Willem van Straten for modifying the software correlator \texttt{SFXC}, and software package \texttt{DSPSR}, respectively, to fit our needs. DH is supported by the Women In Science Excel (WISE) programme of the Netherlands Organisation for Scientific Research (NWO).

%%%%%%%%%%%%%%%%%%%%%%%%%%%%%%%%%%%%%%%%%%%%%%%%%%
\section*{Data Availability}

The relevant code and data products for this work will be uploaded on Zenodo at the time of acceptance for publication.



\bibliographystyle{mnras}
\bibliography{frb} % if your bibtex file is called example.bib

%\begin{thebibliography}{99}
%\end{thebibliography}

%%%%%%%%%%%%%%%%%%%%%%%%%%%%%%%%%%%%%%%%%%%%%%%%%%

%%%%%%%%%%%%%%%%% APPENDICES %%%%%%%%%%%%%%%%%%%%%

\appendix

\section{DM determination}

\begin{figure*}
    \centering
    \subfloat[][]{\includegraphics[width=.475\textwidth]{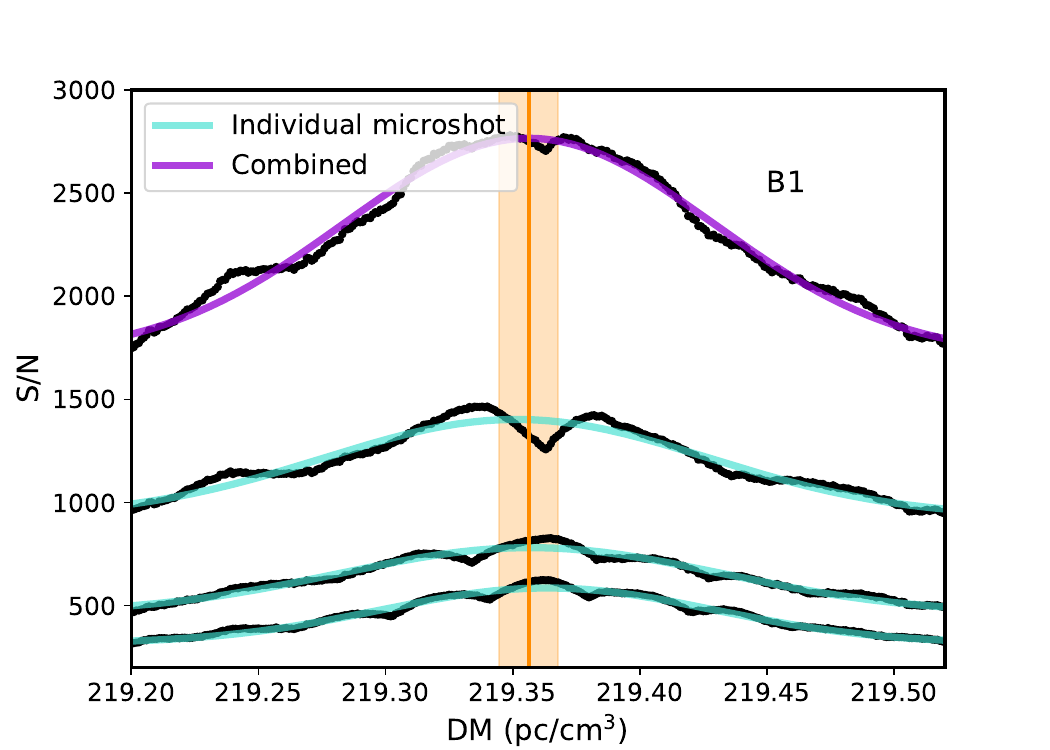}}\quad
    \subfloat[][]{\includegraphics[width=.475\textwidth]{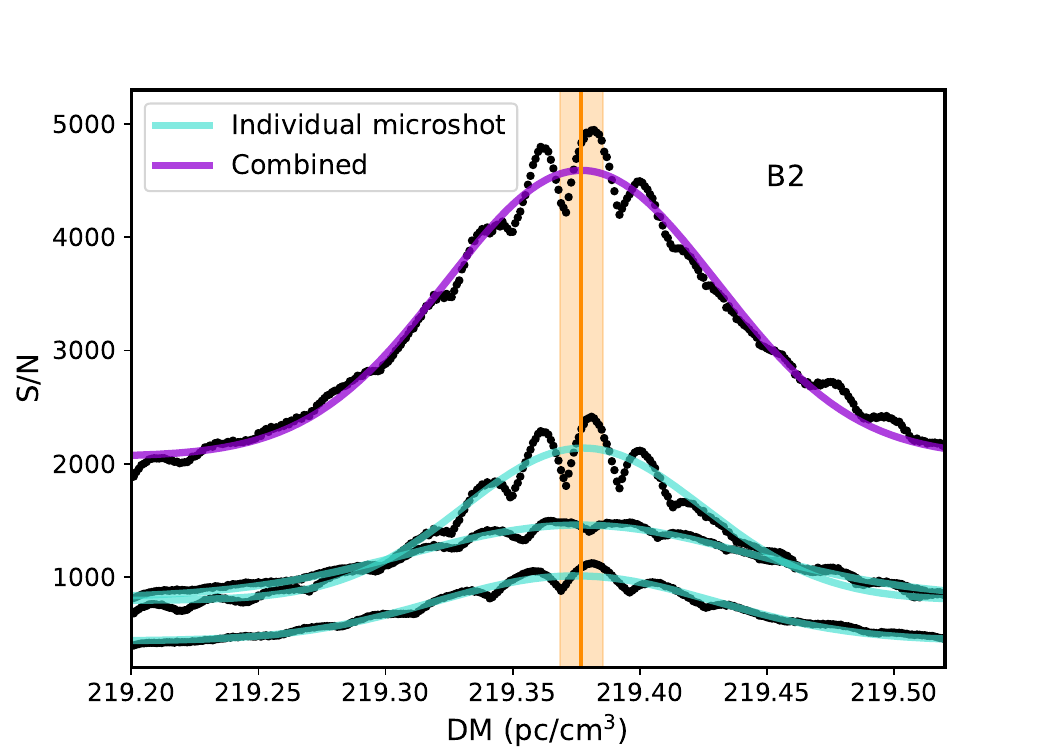}}
    \caption{The S/N as a function of DM of the three brightest microshots in B1 and B2 are shown as black dots. Gaussian fits to these curves are overplotted in turquoise. The combined S/N of the three microshots is also shown, with a Gaussian fit in purple. The centroid of this Gaussian fit is indicated by the solid vertical orange line, while the shaded region shows the 3$\sigma$ uncertainty. The scalloping in the curves is likely the result of the strongly scintillating microshots that have a duration comparable to the time resolution of our data.}
    \label{fig:sn_v_dm}
\end{figure*}

In order to determine the DM we dedispersed the three brightest microshots in B1 and B2 to a range of DMs in the interval $219.2 - 219.55$\,pc\,cm$^{-3}$. The DM is then given by the centroid of a Gaussian fit to the sum of the S/N vs DM curves. These curves are shown in Fiure~\ref{fig:sn_v_dm}. See the main text for more detail.

\section{ACF analysis}
\begin{figure*}
    \centering
  \subfloat[][]{\includegraphics[width=.32\textwidth]{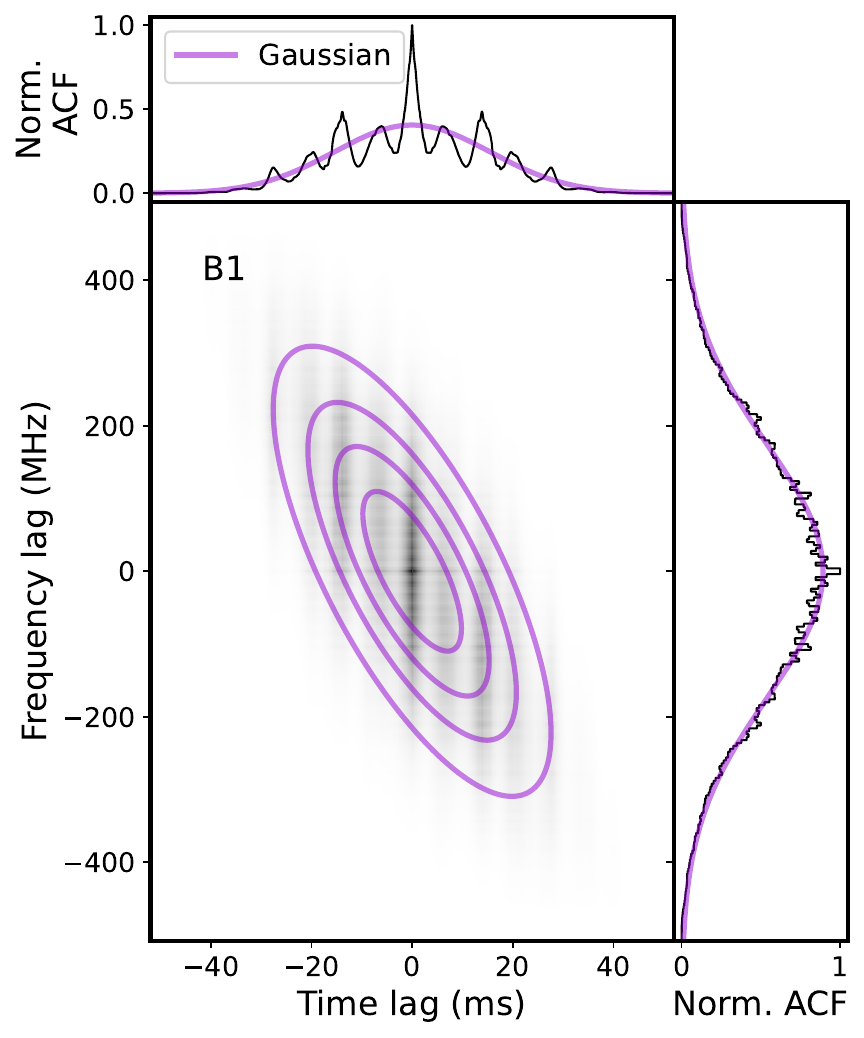}}\quad
  \subfloat[][]{\includegraphics[width=.32\textwidth]{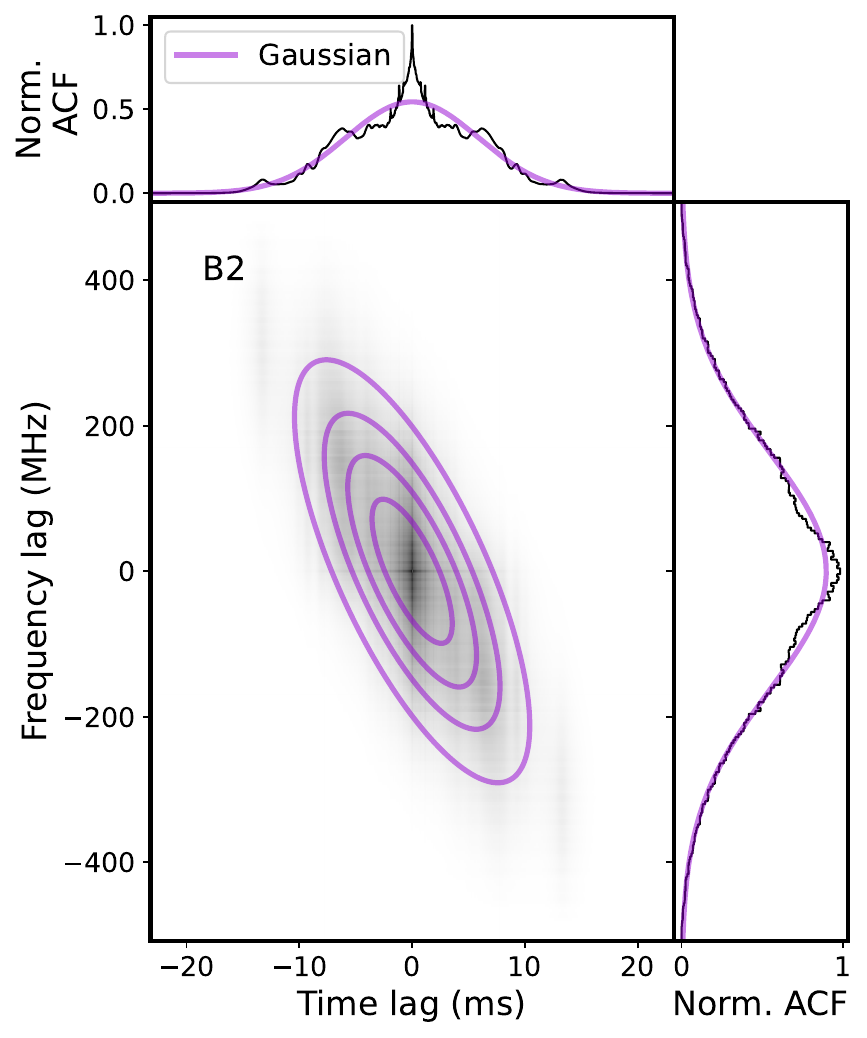}}\quad
  \subfloat[][]{\includegraphics[width=.32\textwidth]{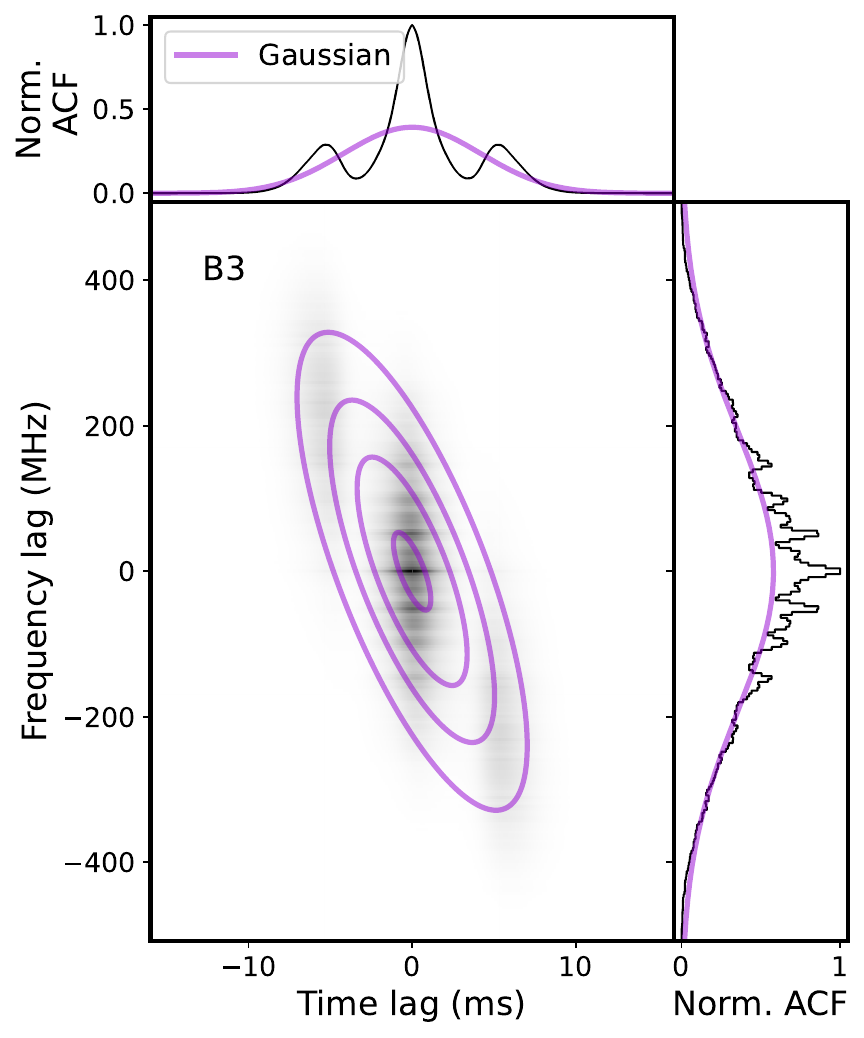}}\quad
  \caption{2D autocorrelation functions of burst B1, B2, and B3 detected with the NRT. A 2D Gaussian is fitted to the 2D ACFs to determine the drift rate. The top panels show the 1D time ACF, while the right hand panels show the 1D frequency ACF. For more detail see the main text. }
   \label{fig:nrt_acf}
\end{figure*}

\begin{figure*}
    \centering
    \includegraphics[width=1\textwidth]{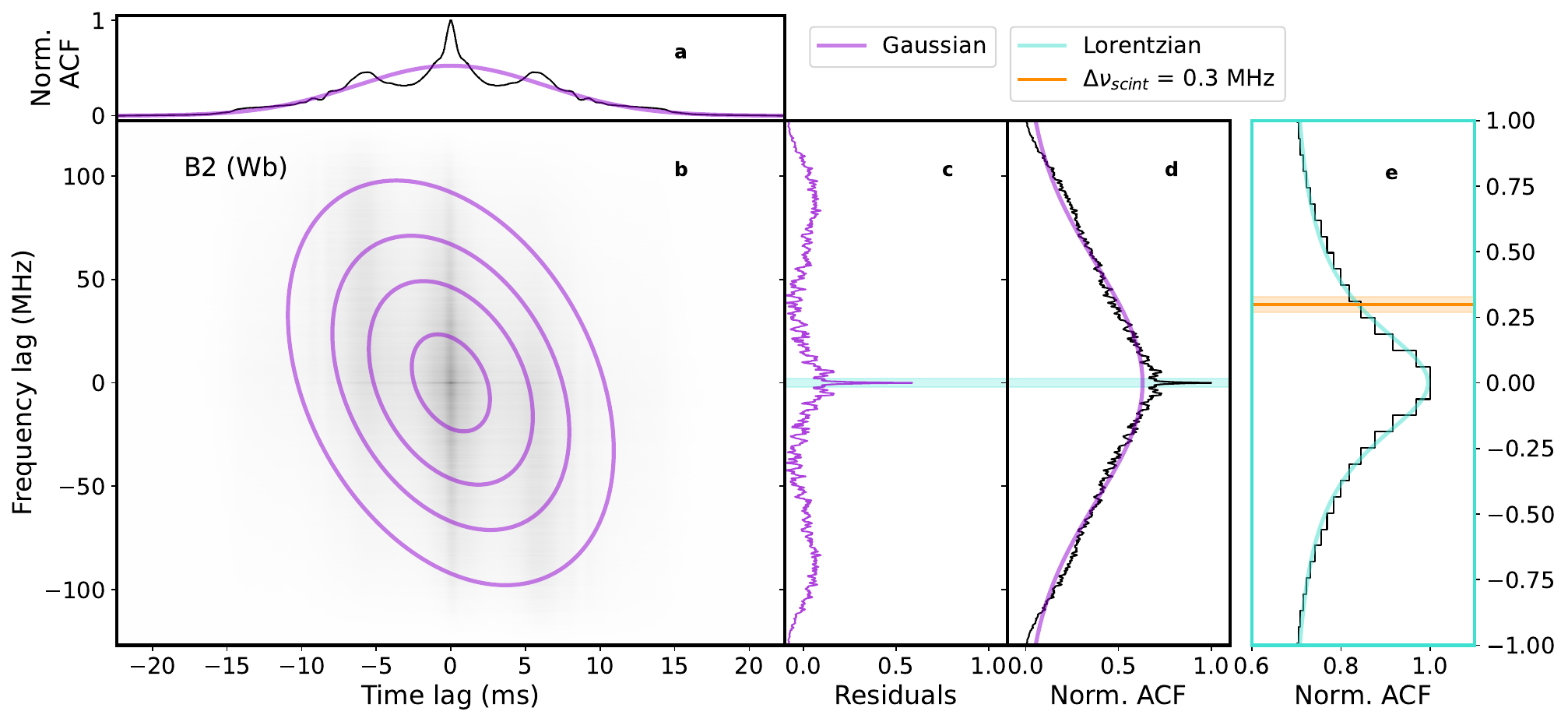}
    \caption{Panels a and c show 1D time and frequency ACFs, respectively, of B2 using the data from the Westebork RT-1 dish. Gaussian fits have been overplotted in purple. Panel b shows the 2D ACF, with a 2D Gaussian fit overplotted in purple. The residuals of the frequency ACF as a fraction of the normalised maximum of the 1D ACF are shown in panel c. In panels c and d, a strong peak in the frequency ACF corresponding to the scintillation bandwidth has been highlighted in turquoise. Finally, we zoom in on this peak in panel e, where a Lorentzian fit is overplotted in turquoise. The HWHM of the Lorentzian function, which corresponds to the scintillation bandwidth, is indicated by the orange line, with the shaded orange regions showing the 3$\sigma$ uncertainty. This value is consistent within a factor of 2 of the NE2001 prediction.  }
   \label{fig:wb_acf}
\end{figure*}

In Figures~\ref{fig:nrt_acf} and \ref{fig:wb_acf} we show 2D ACFs of the bursts presented in this paper. See the main text and figure captions for more detail.

\section{Polarisation Calibration}

\begin{figure*}
    \centering
    \includegraphics[width=\textwidth]{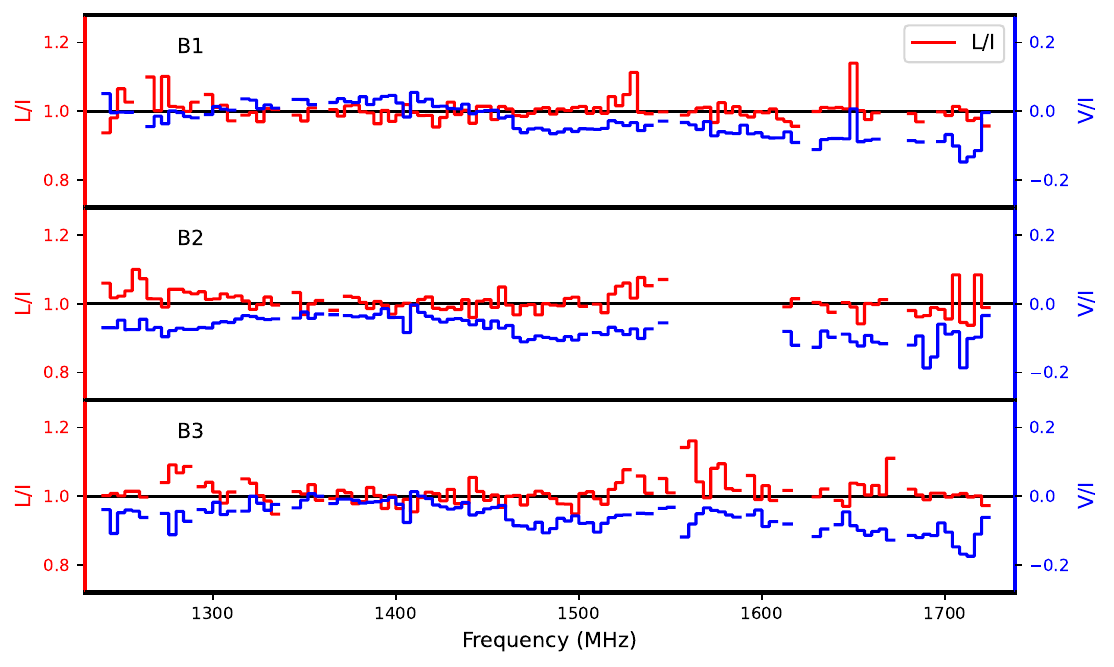}
    \caption{The linear (in red) and circular (in blue) polarisation fractions of B1, B2 and B3 as a function of frequency. Gaps occur due to channels that have been zapped to remove RFI contamination.  }
    \label{fig:stokes_spectra}
\end{figure*}

In Figure~\ref{fig:stokes_spectra} we plot the polarisation fractions of our bursts as a function of frequency. A similar trend in Stokes~V/I for all three bursts indicates a slight imperfection in the polarimetric calibration that does not, however, significantly influence our results or interpretation thereof. 

%%%%%%%%%%%%%%%%%%%%%%%%%%%%%%%%%%%%%%%%%%%%%%%%%%

% Don't change these lines
\bsp	% typesetting comment
\label{lastpage}
\end{document}